\documentclass{aa}  
\usepackage{deluxetable}
\usepackage{graphicx}
\usepackage{longtable}
\usepackage{txfonts}

\usepackage{natbib}
\usepackage[labelsep=period]{caption}
\newcommand\teff{T$_{\rm eff}$}
\newcommand\logg{log~g}
\newcommand\vt{v$_{\rm t}$}

\newcommand\afe{[$\alpha$/Fe] \,}

\begin{document} 

   \title{VLT/FLAMES high-resolution chemical abundances in Sculptor: \\a textbook
dwarf spheroidal galaxy\thanks{Based on VLT/FLAMES observations collected at the European Organisation for Astronomical Research (ESO) in the Southern Hemisphere under programmes 71.B-0641 and 171.B-0588.}$^,$\thanks{Tables C.1-C.5 are only available in electronic form at the CDS via anonymous ftp to cdsarc.u-strasbg.fr (130.79.128.5)
or via \text{http://cdsweb.u-strasbg.fr/cgi-bin/qcat?J/A+A/.}}
}

   \author{V. Hill \inst{1},
   		\'{A}. Sk\'{u}lad\'{o}ttir \inst{2},	
   		E. Tolstoy \inst{3},	
        K.A. Venn \inst{4},	
        M.D. Shetrone \inst{5},	
        P. Jablonka \inst{6,7},	
        F. Primas \inst{8},		
       G.~Battaglia \inst{9,10},	
        T.J.L.~de Boer\inst{11},	
        P. Fran\c cois \inst{7},	
        A. Helmi \inst{3},	
        A. Kaufer \inst{12},	
        B. Letarte	\inst{13},	
        E. Starkenburg\inst{14},	
        \and
        M. Spite\inst{7},		
        }

\institute{Universit\'e C\^ote d'Azur, Observatoire de la C\^ote d'Azur, CNRS, Laboratoire Lagrange, Bd de l'Observatoire, CS~34229, 06304 Nice cedex~4, France
	 \email{Vanessa.Hill@oca.eu}
\and 		
	Max-Planck-Institut f$\ddot{\text{u}}$r Astronomie, K$\ddot{\text{o}}$nigstuhl 17, D-69117 Heidelberg, Germany
\and        
	Kapteyn Astronomical Institute, University of Groningen, PO Box 800, 9700AV Groningen, the Netherlands
\and        
	Department of Physics and Astronomy, University of Victoria, 3800 Finnerty Road, Victoria, BC V8P 1A1, Canada 
\and        
	McDonald Observatory, University of Texas at Austin, Fort David, TX, USA
\and        
	Laboratoire d'Astrophysique de l'\'ecole Polytechnique F\'ed\'erale de Lausanne (EPFL), 1290 Sauverny, Switzerland 
\and
	GEPI, Observatoire de Paris, CNRS UMR 8111, Universit\'e Paris Diderot, 92125, Meudon Cedex, France
\and
	European Southern Observatory, Schwarzschild-Str. 2, 85748 Garching, Germany
\and
Instituto de Astrofisica de Canarias, calle Via Lactea s/n, E-38205 La Laguna, Tenerife, Spain
\and
	Universidad de La Laguna, Dpto. Astrofisica, E-38206 La Laguna, Tenerife, Spain 
\and
	Department of Physics, University of Surrey, Guildford, GU2 7XH, UK         
\and        
	European Southern Observatory, Alonso de Cordova 3107, Vitacura, Casilla 19001, Santiago, Chile
\and
	Centre for Space Research, North-West University, Potchefstroom, 2520, South Africa
\and
	Leibniz Institute for Astrophyics Potsdam (AIP), An der Sternwarte 16, D-14482 Potsdam, Germany
}

  \abstract
{We present detailed chemical abundances for 99 red-giant branch stars in the centre of the Sculptor dwarf spheroidal galaxy, which have been obtained from high-resolution VLT/FLAMES spectroscopy. The abundances of Li, Na, $\alpha$-elements (O, Mg, Si, Ca Ti), iron-peak elements (Sc, Cr, Fe, Co, Ni, Zn), and $r$- and $s$-process elements (Ba, La, Nd, Eu) were all derived using stellar atmosphere models and semi-automated analysis techniques. The iron abundances populate the whole metallicity distribution of the galaxy with the exception of the very low metallicity tail, $-2.3 \leq$ [Fe/H] $\leq -0.9$. There is a marked decrease in [$\alpha$/Fe] over our sample, from the Galactic halo plateau value at low [Fe/H] and then, after a `knee', a decrease to sub-solar [$\alpha$/Fe] at high [Fe/H]. This is consistent with products of core-collapse supernovae dominating at early times, followed by the onset of supernovae type~Ia as early as $\sim$12~Gyr ago. The $s$-process products from low-mass AGB stars also participate in the chemical evolution of Sculptor on a timescale comparable to that of supernovae type~Ia. However, the $r$-process is consistent with having no time delay relative to core-collapse supernovae, at least at the later stages of the chemical evolution in Sculptor. Using the simple and well-behaved chemical evolution of Sculptor, we further derive empirical constraints on the relative importance of massive stars and supernovae type~Ia to the nucleosynthesis of individual iron-peak and $\alpha$-elements. The most important contribution of supernovae type~Ia is to the iron-peak elements: Fe, Cr, and Mn. There is, however, also a modest but non-negligible contribution to both the heavier $\alpha$-elements: S, Ca and Ti, and some of the iron-peak elements: Sc and Co. We see only a very small or no contribution to O, Mg, Ni, and Zn from supernovae type~Ia in Sculptor. The observed chemical abundances in Sculptor show no evidence of a significantly different initial mass function, compared to that of the Milky Way. With the exception of neutron-capture elements at low [Fe/H], the scatter around mean trends in Sculptor for $\text{[Fe/H]}>-2.3$ is extremely low, and compatible with observational errors. Combined with the small scatter in the age-elemental abundances relation, this calls for an efficient mixing of metals in the gas in the centre of Sculptor since $\sim$12~Gyr ago. 
}


\authorrunning{Hill et al.}
\titlerunning{High-resolution chemical abundances in Sculptor}

\maketitle
%

\section{Introduction}

Measuring the detailed abundances of a variety of chemical elements in individual stars in a galaxy is the most accurate way to trace the chemical evolution processes through time.  The chemical abundance pattern of each star is the product of the enrichment caused by all the previous generations of stars \citep[e.g.][]{Tinsley:1979,Tinsley81,MatteucciFrancois89,McWilliam:1997}.  In the Local Group we are in the unique position to be able to study a wide range of galaxies in extraordinary detail, star by star. The signatures of different physical processes allow us to disentangle the star formation and evolutionary properties of nearby galaxies back to the earliest times.

The Sculptor dwarf spheroidal (dSph) galaxy is a satellite of the Milky Way at a distance of $86\pm 5$~kpc \citep{Pietrzynskietal:2008}, and at high Galactic latitude ($b=-83$~degrees), with a systemic velocity, $v_{hel} = 110.6 \pm 0.5$~km$\,$s$^{-1}$ \citep{Quelozetal:1995, Battaglia08a}.  This makes it a relatively straightforward target for detailed studies of its resolved stellar population, as it is close enough for its red-giant branch (RGB) stars to be targeted with high-resolution (HR) spectroscopy. There is little Galactic foreground contamination, most of which can be easily distinguished by velocity and a careful analysis of the spectra \citep[e.g.][]{BattagliaStarkenburg12}.  In contrast to the smaller ultra-faint dwarf (UFD) galaxies, the number of bright RGB stars that can be studied individually in a dSph is significantly larger, making the conclusions based on the properties of the resolved stellar population less prone to the effects of small number statistics.  

There have been numerous photometric studies of the resolved stellar population in Sculptor since its discovery by Shapley in the
1930s, e.g. \citet{Hodge65,  NorrisBessell78, Kaluzny95, Monk99, Hurley-Kelleretal:1999,Majewski99, Harbeck01, Dolphin02, Maccarone05, Babusiaux05, Westfall06, Mapelli09, Menzies11, deBoer11, deBoer12, Salaris13, MV15,  MV16,Savino18}. This includes colour-magnitude diagram (CMD) analyses, but also the study of individual populations, such as the horizontal branch, X-ray binaries, blue stragglers and asymptotic giant branch (AGB) stars. The star formation history, coming from a careful CMD analysis, shows a peak in star formation $\sim$12~Gyr ago, with a subsequent tail-off in the star formation rate \citep{deBoer12}, until Sculptor stopped forming stars $\sim$8~Gyr ago \citep[e.g.][]{Hurley-Kelleretal:1999, Dolphin02,deBoer12}. At the present time, Sculptor does not have any associated \ion{H}{I}~gas \citep{GrcevichPutman:2009}. By combining CMD analysis with the spectroscopically determined metallicities for individual stars, \citet{deBoer12} determined ages for the RGB stars in Sculptor. This made it possible for the first time to put accurate timescales on the chemical evolution processes in a dSph galaxy.

Early kinematic studies established that the Sculptor dSph is dominated by dark matter \citep{Dacosta91, Queloz95, Aaronson87, Tolstoy01}. The total mass of Sculptor is $(3.4\pm 0.7 )\times 10^8$M$_{\odot}$, which represents a mass-to-light ratio of $158\pm 33~$(M/L)$_{\odot}$ inside 1.8~kpc, with tentative evidence for a velocity gradient of $7.6^{+3.0}_{-2.2}$~km$\,$s$^{-1}\,$deg$^{-1}$ \citep{Battaglia08a}. This gradient can be interpreted as rotation about the minor axis, or it could be due to tidal disruption by the Milky Way. The combination of Hubble Space Telescope (HST) and Gaia observations of individual stars in Sculptor \citep{Massari17} has provided a new and accurate proper motion and orbit determination for Sculptor, which was further refined by Gaia DR2 results \citep{Helmi18}, see Table~\ref{tab:gaia}. These new determinations are fairly different from previous estimates in the literature \citep{Schweitzer95, Piatek06, Walker08, Sohn17}. In this relatively small and simple galaxy there are two distinct stellar populations present. They have different kinematics, metallicity, and spatial distributions \citep{Tolstoy04, Helmi06, Colemanetal:2005, Clementini05, Battaglia08a}, with one population that is centrally concentrated, kinematically cold and relatively metal-rich; and another that is a more spatially extended, kinematically warmer, and more metal-poor.  

\begin{table}
\caption{Astrometry of the Sculptor dwarf spheroidal galaxy by the \citet{Helmi18}: the position on the sky ($\alpha$,$\delta$), parallax $\varpi$, proper motions ($\mu_{\alpha^*}$,$\mu_\delta$),  and the elements of the covariance matrix, $\epsilon_X$. Also included are the number of member stars, $N_\star$ as determined by Gaia for the magnitude limit,~$G_\textsl{lim}$.
}
\label{tab:gaia}
\centering
\small
\begin{tabular}{l r l}
\hline\hline
\multicolumn{3}{c}{The Sculptor dSph}\\
\hline
$\alpha$	 & 15.0392  & deg\\
$\delta$	&	 $-33.7092$ & deg\\
$\varpi$ &	$-0.013$		& mas\\
$\epsilon_\varpi$ & 0.004	&	mas\\
$\mu_{\alpha^*}$ & 0.082	&	mas/yr\\
$\epsilon_{\mu_{\alpha^*}}$& 0.005	&	mas/yr\\
$\mu_{\delta}$ & -0.131& mas/yr\\
$\epsilon_{\mu_{\delta}}$ & 0.004& mas/yr\\
$G_{lim}$ &19.5 & mag\\
$N_\star$ & 1592& \\
\hline
\end{tabular}
\end{table}

The first detailed analysis of chemical abundances in Sculptor stars came from VLT/UVES spectra \citep{Shetrone03, Tolstoy03, Geisler05}, examining 9 individual RGB stars in total. The position of the knee in the $\alpha$-elements was found to be at a significantly lower [Fe/H] than any other stellar system previously measured \citep{Tolstoy03, Venn04}.  This sample of 9 stars, however, was too small to make concrete general conclusions, especially about the degree of scatter in the abundances. An extensive intermediate-resolution spectroscopic survey with Keck/Deimos of nearly 400 RGB stars around the centre of the Sculptor dSph determined the abundances of Fe, Mg, Ca, Si and Ti, using the synthesis of a large numbers of weak lines over a large wavelength range \citep{Kirby11}. Other studies have focused on one or more individual stars \citep[e.g.][]{ Smith83, Shetrone98, Salgado16, Skuladottir15b}, or individual elements, such as Mn \citep{North12}. Recently, S and Zn were also measured in Sculptor \citep{Skuladottir15b, Skuladottir17}, and then compared directly to chemical abundances observed in damped Lyman-$\alpha$ systems observed at high redshifts \citep{Skuladottir18}. 

Sculptor has also been the target of extensive searches for extremely metal-poor stars \citep{Kirby11,Starkenburg10,Chiti18}, feeding high-resolution follow-ups to verify the detailed chemical abundances of this elusive population \citep{Tafelmeyer10,Frebel10,Starkenburg13,Jablonka15, Simon15, Chiti18}.  Among these, the most metal-poor star outside the Milky Way was found at $\text{[Fe/H]}=-3.96\pm0.06$  \citep{Tafelmeyer10}. The metal-poor tail of the Sculptor dSph shows both similarities and differences with their counterparts in the Galactic halo.  

In particular, the Milky Way halo stars show a bimodality in carbon (e.g. \citealt{Aoki07, Placco14} and references therein), with two separated populations, above $\text{[C/Fe]}=0.7$ (CEMP stars), and below (C-normal stars). Among these, CEMP-no stars (with no enhancement in neutron-capture elements Ba or Eu abundances) are believed to show chemical signatures of the very first stars (e.g. \citealt{UmedaNomoto02,Meynet06}). Carbon has been measured in sizeable samples of RGB stars in the Sculptor dSph using low-resolution (LR) spectroscopy: with Keck/Deimos by \citet{Kirby15}; VLT/VIMOS by \citet{Lardo16}, also including nitrogen; and with Magellan-Clay/M2FS by \citet{Chiti18}. Neither the HR surveys of extremely metal-poor stars \citep{Tafelmeyer10,Frebel10,Starkenburg13,Jablonka15, Simon15}, nor the earlier LR studies \citep{Kirby15,Lardo16} found any CEMP-no stars in Sculptor. However, one CEMP-no star was found at a surprisingly high $\text{[Fe/H]}=-2$ \citep{Skuladottir15a}, showing clear differences in [C/Fe] compared to other stars at this metallicity in Sculptor. 

The recent study of \citet{Chiti18}, focusing on the most metal-poor tail in Sculptor ($\text{[Fe/H]}\leq-3$) with LR spectroscopy ($R\sim2000$), found a trend of increasing [C/Fe] towards the lowest metallicities, as predicted in \citet{Salvadori15}. Their measured fraction of CEMP-no stars was 24\%  at $\text{[Fe/H]}\leq-3$, which is consistent with that observed in the Milky Way halo, $\sim$43\% \citep{Placco14}, given their errors. However, no CEMP-no stars were measured to have $\text{[C/Fe]}>+1$ in Sculptor, while the fraction of such stars in the Milky Way halo is $\sim$32\% at $\text{[Fe/H]}\leq-3$ \citep{Placco14}. 

\begin{table}
\caption{\label{table:log} Observing log, as well as the grating setting used for each spectrograph, the plate or fibre set used, the exposure time (Expt), the airmass (AirM) and when available the seeing measurement from the seeing monitor (DIMM).
}
\centering
\begin{tabular}{llllcc}
\hline\hline
Date & Setting & Plate & Expt (s) & AirM  & DIMM \\
\hline
2003-08-24 &  HR10  & MED1 & 3600 & 1.02 & - \\
2003-08-24 &  HR10  & MED1 & 3600 & 1.02 & 0.99 \\
2003-08-28 &  HR10  & MED2 & 3600 & 1.03 & - \\
2003-08-25 &  HR10  & MED2 & 5400 & 1.03 & 0.81 \\ 
2003-08-22 &  HR13  & MED1 & 3600 & 1.01 & 0.70 \\
2003-08-22 &  HR13  & MED1 & 3600 & 1.07 & 0.67 \\
2003-08-20 &  HR13  & MED2 & 4200 & 1.00 & 0.98 \\ 
2003-08-20 &  HR13  & MED2 & 4200 & 1.01 & 0.84 \\
2003-08-21 &  HR14A & MED2 & 3600 & 1.01 & 0.66 \\
2003-08-21 &  HR14A & MED2 & 3600 & 1.04 & 0.78 \\
2003-08-21 &  HR14A & MED2 & 3600 & 1.14 & 0.66 \\
2003-08-22 &  HR14A & MED2 & 4500 & 1.05 & 1.00 \\
2003-08-21 &  HR14A & MED2 & 4700 & 1.04 & 0.84 \\
2003-08-23 &  HR14A & MED2 & 5400 & 1.04 & 1.06 \\
2003-08-23 &  HR15  & MED1 & 3600 & 1.01 & 0.65 \\
2003-08-23 &  HR15  & MED1 & 3600 & 1.06 & 0.76 \\
\\
2003-08-23 &  580 & FIB1 & 3600 & 1.01 & 0.65 \\
2003-08-22 &  580 & FIB1 & 3600 & 1.01 & 0.70 \\
2003-08-23 &  580 & FIB1 & 3600 & 1.06 & 0.76 \\
2003-08-22 &  580 & FIB1 & 3600 & 1.07 & 0.67 \\
2003-08-20 &  580 & FIB2 & 4200 & 1.00 & 0.00 \\ 
2003-08-20 &  580 & FIB2 & 4200 & 1.01 & 0.84 \\
2003-08-24 &  580 & FIB1 & 3600 & 1.02 & - \\
2003-08-24 &  580 & FIB1 & 3600 & 1.02 & 0.99 \\
2003-08-28 &  580 & FIB2 & 3600 & 1.03 & - \\
2003-08-22 &  580 & FIB2 & 4500 & 1.05 & 1.00 \\
2003-08-21 &  580 & FIB2 & 4700 & 1.04 & 0.84 \\
2003-08-21 &  580 & FIB2 & 5400 & 1.02 & 0.69 \\
2003-08-25 &  580 & FIB2 & 5400 & 1.03 & 0.81 \\ 
2003-08-23 &  580 & FIB2 & 5400 & 1.04 & 1.06 \\
2003-08-21 &  580 & FIB2 & 5400 & 1.14 & 0.66 \\
\hline
\end{tabular}
\end{table}


Given the available large and detailed spectroscopic and photometric surveys of its stellar population, the Sculptor dSph is an obvious template for understanding galaxy formation and evolution on small scales. This galaxy has therefore also been the target of a large number of dedicated modelling efforts, using different techniques and approaches, e.g. \citet{LanMatt03, LanMatt04, Fenner06, Kawata06, Salvadori08, Marco08, Revazetal:2009, RevazJablonka12, RevazJablonka18, RomanoStarkenburg13, Vinc16, Cote17}. 

Here we present HR spectra for 99~RGB stars in this galaxy taken with ESO VLT/FLAMES as part of the DART survey \citep{Tolstoy06}. This study has been presented (without any technical details) in \citet{Tolstoy09}, and has already been used in a number of other publications. With the same spectra and stellar parameters as used here, \citet{North12} measured Mn abundances in Sculptor and investigated its nucleosynthetic origin. The stellar parameters determined here have also been used in the study of S and Zn in this galaxy \citep{Skuladottir15a,Skuladottir17}. In addition, these results have been used in the verification of the \ion{Ca}{II} triplet metallicity scale \citep{Battaglia08b, Starkenburg10}. The [Fe/H] and [$\alpha$/Fe] abundances were used in the CMD analysis in determining the star formation history in Sculptor \citep{deBoer12}. The data presented here has also been used extensively as constraints for models, by \citet{Revazetal:2009, RevazJablonka12, RevazJablonka18, RomanoStarkenburg13, Cote17}. 

Combining all the available results, it is clear that there are significant differences in the chemical abundances of Sculptor and the Milky Way, both at high and low metallicities. Detailed chemical abundances in Sculptor, such as those presented here, are therefore necessary to help us better understand this intriguing galaxy.


\section{Data collection and pipeline processing}

As part of the Paris Observatory VLT/FLAMES Guaranteed Time Observations (GTO) allocation, we carried out a spectroscopy programme of individual RGB stars over a 25$\arcmin$ diameter field of view at the centre of  the Sculptor dSph galaxy. We simultaneously used FLAMES/GIRAFFE, in HR Medusa mode, and the fibre feed to the FLAMES/UVES spectrograph on VLT UT2 \citep{Pasquinietal:2002}.  These observations were carried out between 20-28 August 2003. In Table~\ref{table:log} the details of the observations are given.

\subsection{Sample selection}

\begin{figure}
\centering
\includegraphics[width=\hsize-0.5cm,angle=0,clip=]{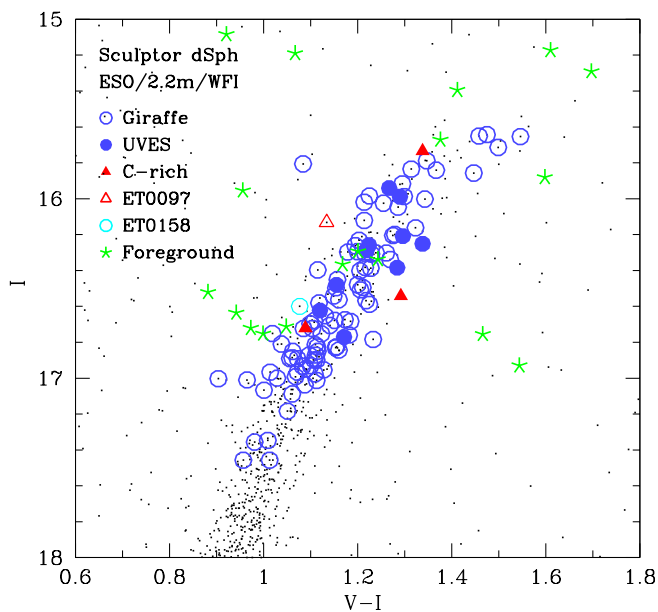}
\caption{ESO/2.2m/WFI photometry (I, (V-I)) CMD of the central 30$\arcmin$ of Sculptor. Our spectroscopic target selection is overlaid. Foreground contamination stars are green asterisks and other symbols denote Sculptor members: blue circles for the main UVES (filled) and GIRAFFE (open) samples. The Li-rich star ET0158 is shown as a cyan open circle. Red filled triangles show C-rich stars and the CEMP-no star, ET0097, is shown as a red open triangle.
}
\label{fig:cmd} 
\end{figure}

\begin{figure}
\centering
\includegraphics[width=\hsize-0.5cm,angle=0,clip=]{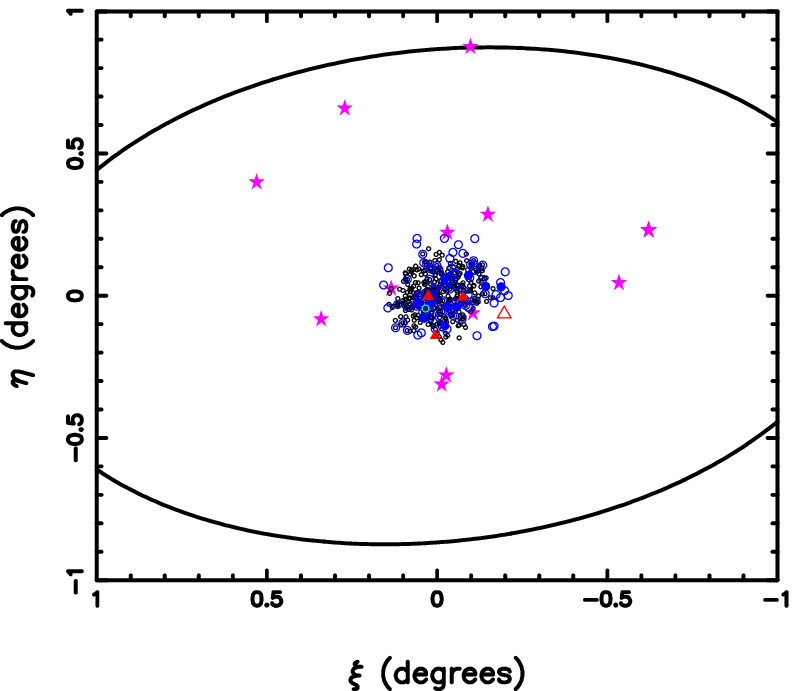}
\caption{Spatial scale of Sculptor with tidal radius (black ellipse). Symbols are the same as in Fig.~\ref{fig:cmd}, with the addition of black open circles as LR data \citep{Kirby11} and magenta stars as HR data from the literature \citep{Shetrone03,Geisler05,Frebel10,Starkenburg13,Simon15,Jablonka15}.
}
\label{fig:sculptor} 
\end{figure}

Our target RGB stars were randomly selected within the FLAMES field of view from the I, (V-I) CMD shown in Fig.~\ref{fig:cmd}. The spatial scale of the targets are shown in Fig.~\ref{fig:sculptor}. We limited ourselves to the upper part of the RGB, with I $<$ 17.5, to maximise the signal-to-noise (S/N). From the 132 fibres available in the Medusa mode of FLAMES/GIRAFE we allocated 117 to known and likely RGB stars in the Sculptor dSph, and 15 to monitor the sky background.  For FLAMES/UVES, 6 fibres were allocated to RGB stars in Sculptor and 2 to the sky. The UVES configuration was changed once in the course of our FLAMES/GIRAFFE observations to give a total of 12 stars observed with FLAMES/UVES.

\begin{table}
\caption{Wavelength range, resolution and observing time of the GIRAFFE and UVES settings used here.
}
\label{tab:FLAMES}
\centering
\small
\begin{tabular}{l c c c c}
\hline\hline
Setting & $\lambda_\textit{min}$ & $\lambda_\text{max}$ & Resolution & Obs. time \\
& [$\AA$] & [$\AA$] & &\\
\hline
HR10	 & 5340 & 5620 & 19\,800 & 4hr30min\\
HR13	 & 6120 & 6400 & 22\,500 & 4hr20min\\
HR14A	 & 6390 & 6620 & 28\,800 & 7hr\\
HR15	 & 6610 & 6960 & 19\,300 & 2hr\\
UVES     & 4800 & 6800 & 47\,000 & 7 and 11hr\\
\hline
\end{tabular}
\end{table}

\subsection{GIRAFFE and UVES fibre observations}

For the FLAMES/GIRAFFE observations, one Medusa fibre configuration was used for four different wavelength
regions (or settings), chosen to optimise the number \ion{Fe}{I} and \ion{Fe}{II} absorption lines and to observe specific $\alpha$-elements, iron-peak and heavy elements. The total observing time was $\sim$18\,hours divided between 4 HR GIRAFFE settings: HR10, HR13, HR14A, and HR15, see Table~\ref{tab:FLAMES}. The resolution of these different settings ranges from $R\sim19\,000 - 29\,000$. 

The FLAMES/UVES fibres were fed into the red arm of UVES, centred at 580~nm, where the 1$\arcsec$ fibres yield a resolution $R\sim47\,000$ over the wavelength range, see Table~\ref{tab:FLAMES}. Two FLAMES/UVES fibre configurations were used and one contained brighter targets than the other, and so the total exposure time spent on the six brighter and the six fainter targets amounted to 7~hr and 11~hr, respectively.

\subsection{Pipeline reduction}

The FLAMES/GIRAFFE spectra were reduced, extracted and wavelength calibrated using the GIRBLDRS pipeline provided by the FLAMES consortium (written by A.~Blecha and G.~Simon at Geneva Observatory). Each target spectrum was automatically continuum-corrected and cross-correlated with a spectral mask before being coadded. Various sky-subtraction schemes were tested, and there was a negligible difference between them for these HR spectra. We used the same sky-subtraction method as we have used on low-resolution \ion{Ca}{II} triplet observations of Sculptor giants \citep{Battaglia08a} written by M.~Irwin, which scales the sky background to be subtracted from each object spectrum to match the observed sky emission lines.  

The radial velocities were measured by cross-correlating each of the four frames obtained with the HR10 setup spectra against a template (binary mask delivered within the GIRBLDRS pipeline). The measurements are reported in Table~\ref{table:PhotVr}, showing the mean and the dispersion around the mean ($\rm0.7~km\,s^{-1}$ on average) of these individual measurements. A further discussion about the radial velocities in this sample can be found in \citet{Skuladottir17}, where systematic errors between observations taken from 2003-2013 were discussed in more detail. Six stars (ET0094, ET0139, ET0163, ET0173, ET0206, and ET0369) showed significant velocity variations, more than 2$\sigma$ from the median, and two stars (ET0097, and ET0109) showed moderate velocity variations, 1-2$\sigma$ from the median \citep{Skuladottir17}.

For equivalent width (EW) determination, we used DAOspec \citep{StetsonPancino:2008}, which determines EWs from Gaussian fitting for a single FWHM value, determined per target, combined with an iterative fit to the global continuum (examined further by \citealt{Letarte10}). We were able to verify the zero point accuracy of the FLAMES/GIRAFFE observations for both velocity and
EW determinations by deliberately reobserving six stars previously observed with UVES and analysed by \citet{Shetrone03} and \cite{Geisler05}. These UVES spectra have both a broader wavelength coverage and higher resolution than the newer GIRAFFE spectra. They thus provide a calibration of the automated methods used here as well as a comparison to the limited wavelength range of the GIRAFFE spectra.

The FLAMES/UVES fibre spectra were treated similarly to GIRAFFE spectra: they were reduced using the FLUVES pipeline \citep{FLUVESpipe04}, sky-subtracted using the same recipe as for GIRAFFE spectra albeit using a single sky fibre, and radial velocities were obtained by cross-correlating the individual 3 to 13 exposures for each star to a template (the observed spectrum, shifted to rest-frame, of star H497 from \citealt{Shetrone03}). Table~\ref{table:PhotVr} reports the mean and dispersion around the mean of these measurements. For the EW determination from these higher resolution spectra, gaussian fits using a single full width half maximum (as performed by DAOspec) is not adequate for the stronger lines, so that EWs were measured manually with the standard {\em splot} routine in IRAF.

\subsection{Pipeline output and member selection}

From the radial velocities, $v_r$, we determined if each star is a likely member of the Sculptor dSph. We also checked the quality (S/N) of the spectra, and if the star is likely to be an RGB star. The S/N for the GIRAFFE spectra was estimated as the inverse of the residuals reported by DAOspec, averaged over all four setups (HR10, HR13, HR14A and HR15). This is only intended to give an indication of the relative quality of the spectra. The S/N reported for the UVES spectra was estimated in a more traditional way, by assessing the dispersion around the continuum in a line-free region of the spectrum around 6400~\AA.

Stars that are not likely members of Sculptor, are not RGB stars, or have spectra of too low quality are removed from further analysis at this point. Table~\ref{table:PhotVr} lists the entire target list for our observations, including non-members and other stars we could not analyse properly. We include the available photometry, V, I, J, K filters \citep{Babusiauxetal:2005,Battaglia08b}, and the measured radial velocities, v$_r$, from the HR10 grating (see previous section), the final coadded S/N of the spectra and also the cross-IDs of stars previously observed with UVES in slit mode.

From the 117 stars observed with GIRAFFE, 17 were found to be non-members based on their radial velocities. One
additional star (ET0092) was rejected because its spectroscopic gravity showed it to be a foreground Galactic sub-giant, with a radial velocity comparable to that of Sculptor. This was also confirmed by an independent automatic classification \citep{Kordopatisetal:2013}. Six stars were excluded because of low S/N, two of those had $\text{S/N}\leq13$, and other four had low $\text{S/N}\leq25$ combined with low metallicity, $\text{[Fe/H]}<-2$. One GIRAFFE spectrum (star ET0041) 
was severely affected by a CCD defect (a bad column) running right through the centre of the fibre image, and was therefore also discarded. One C-star (ET0167, star number 3 from \citealt{Azzopardi85}) and 2 CN-rich stars (ET0136 and ET0315) were also rejected from further analysis, see Fig.~\ref{fig:cmd}, because of the severe blending created by the forest of CN molecular lines. This left 89 stars in the GIRAFFE sample that could be fully analysed. The 12 FLAMES/UVES target stars were all known to be Sculptor dSph members from previous observations.  Two stars were found to have too low S/N for a reliable analysis, and were excluded, leaving a total of 10 UVES spectra. Thus, the full sample of FLAMES observations for which we could proceed to derive stellar parameters, metallicities and detailed abundance ratios consists of 99 stars (89 from the GIRAFFE and 10 from the UVES samples).

To further our membership analysis based on radial velocities and gravities, we also inspected the Gaia DR2 candidate members for Sculptor \citep{Helmi18} which is based on proper motions and CMD position. All our proposed members are in this catalogue, with the exception of six stars (ET0024, ET0048, ET0109, ET0137, ET0173 and ET0378), which have proper motions compatible with Sculptor but were discarded from the Sculptor members because of their location in the Gaia DR2 (G,(BP-RP)) CMD. Conversely, one star in the \citet{Helmi18} catalogue is discarded as a Sculptor member in the present work based on its radial velocity (ET0124). We are thus confident that the members that we have identified here are indeed members of Sculptor.

\begin{figure*}
\centering
\includegraphics[height=\hsize,angle=-90,clip=]{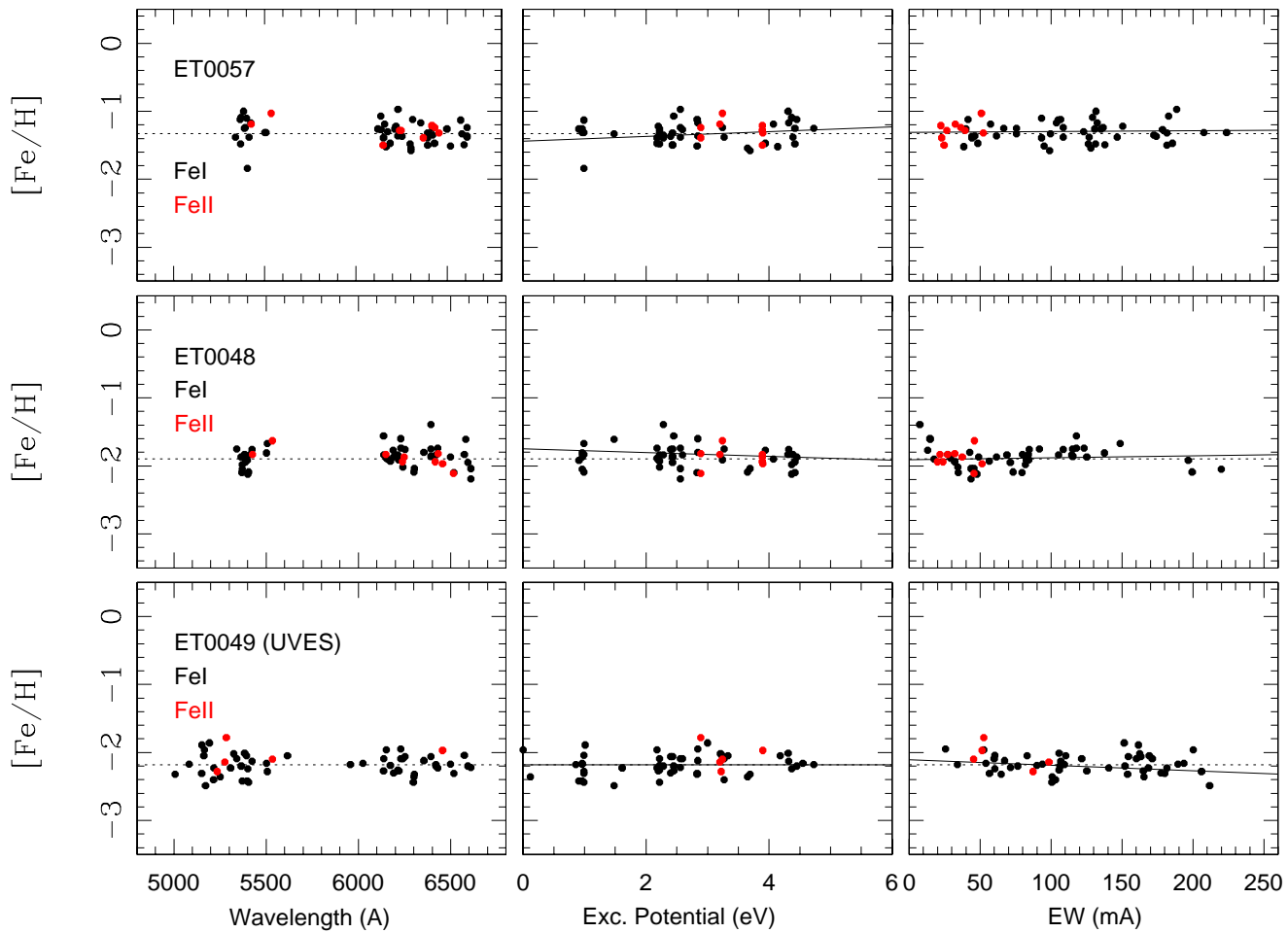}
\caption{Diagnostic plots for three typical stars in our sample. Top row: ET0057, $\text{[Fe/H]}=-1.3$, GIRAFFE. Middle row: ET0048, $\text{[Fe/H]}=-1.9$, GIRAFFE. Bottom row: UET0049, $\text{[Fe/H]}=-2.2$, UVES. For each star, iron abundances (\ion{Fe}{I}: black; \ion{Fe}{II}: red) are plotted against wavelength, excitation potential and EW of the line. Dotted lines are the mean [Fe/H] of each star while solid lines show the slopes of best fits.}
\label{fig_diagnostics1} 
\end{figure*}

\section{Stellar parameters and model atmospheres}

A comprehensive model atmosphere abundance analysis was performed for our sample of 99 stars in Sculptor's central field. The GIRAFFE and UVES spectra were treated separately, because of the difference in spectral resolution and wavelength range, see Table~\ref{tab:FLAMES}. We follow the method outlined in \citet{Shetrone03} and \citet{Venn12} for the UVES spectra, and that outlined by \citet{Letarte10} and \citet{Lemasle12,Lemasle14} for the GIRAFFE spectra, with some minor adjustments to take advantage of the higher signal to noise ratio of the present sample.

\subsection{The line list}

The line list and atomic data (excitation potential, $\chi$, and log$\,gf$) were adopted from \citet{Shetrone03}, with a few
additional lines selected from the work of \citet{Pompeiaetal:2008} in the LMC. The broadening coefficients (C6) were updated from \citet{Barklemetal:2000,Barklemetal:2005}. All the lines were carefully examined using spectral synthesis to ensure there were no significant blends at our metallicity range in Sculptor, given the GIRAFFE spectral resolution. These were also compared to the previously published UVES results \citep{Shetrone03,Geisler05} using the overlapping sample. The continuum level is more difficult to determine at the lower spectral resolution of the GIRAFFE spectra.  In addition, it can be affected by CN molecular features.  Thus, we have been careful to only adopt lines that are not contaminated by these features for the abundance analysis. The resulting list of reliable stellar absorption lines in our spectra, within the GIRAFFE wavelength range (and including additional lines which are used for the UVES spectra) is given in Table~\ref{table:linelist}.

\subsection{Stellar parameters - photometry}\label{sec_photAPs}

The initial estimates for effective temperature, \teff, and surface gravity, \logg, are based on photometry.  The V and I photometry come from the ESO-MPG 2.2~m telecope and the wide field imager, WFI \citep{Battaglia08b}. The J and K$_s$ photometry, available for a sub-sample of the observed stars, come from the Cambridge Infrared Survey Instrument \citep{Babusiauxetal:2005}. The \teff\ of all observed stars were determined using the (V-I), and where possible also (V-J) and (V-K$_s$) temperature calibrations of \citet{RamirezMelendez05}, after global dereddening by E(B-V) $=$ 0.02, with A(V) $=$ E(B-V)$\times$3.24, E(V-I) $=$ E(B-V)$\times$1.28, E(V-K) $=$ E(B-V)$\times$2.87, E(V-J) $=$ E(B-V)$\times$2.335. Initial metallicity estimates used in the colour-temperature calibration were taken from the LR \ion{Ca}{II} triplet survey \citep{Battaglia08a}. The (V-I) and (V-K) colours and temperatures are listed in Table~\ref{table:Stellpar}, which also includes the physical surface gravity based on the bolometric correction from \citet{Alonsoetal:1999}, assuming the photometric temperature and the initial metallicity for each star, to calculate M$_{bol}$. A distance modulus of (M-m)$_{\rm o} = 19.54$ was adopted from \citet{Mateo:1998}, as in \citet{Tolstoy03}, and a mass of 0.8$\,M_{\sun}$ for each star is assumed to be a reasonable hypothesis, given the age range of the sample \citep[see][]{deBoer12}.

We found the temperatures determined from (V-I) to be on average 200~K hotter than those from (V-J) or (V-K$_s$) for the sub-sample of stars that had IR photometry (56 stars out of the total sample of 99). Since the cause of this offset is not clear, the (V-I), (V-J) and (V-K$_s$) temperature results were averaged together. In the case where either (V-I) or the IR colours were missing, an average offset was applied to ensure that all stars are on the same mean temperature scale.

One possibility to explain this inconsistency would be zero point uncertainties in the photometry. When we used the infrared based temperatures alone, (V-J) or (V-K$_s$), the stellar gravities deduced from ionisation equilibrium of \ion{Fe}{i} and \ion{Fe}{ii}, were too low by a large factor ($\Delta\text{\logg}=0.75$~dex) compared to photometric gravities. A simple zero-point shift in the I photometry to bring (V-I) temperatures in line with infrared ones would therefore result in a temperature scale producing a very uncomfortable ionisation balance. Conversely, shifting the infrared colours to the (V-I) temperature scale required an unreasonably large zero-point offset in the K and J-band photometry.  The solution adopted here (averaging the temperature from three colour indices) produces a temperature scale in good agreement with excitation and ionisation equilibria of the iron lines, and was therefore preferred.

\begin{figure}
\centering
\includegraphics[height=\hsize,angle=0,clip=]{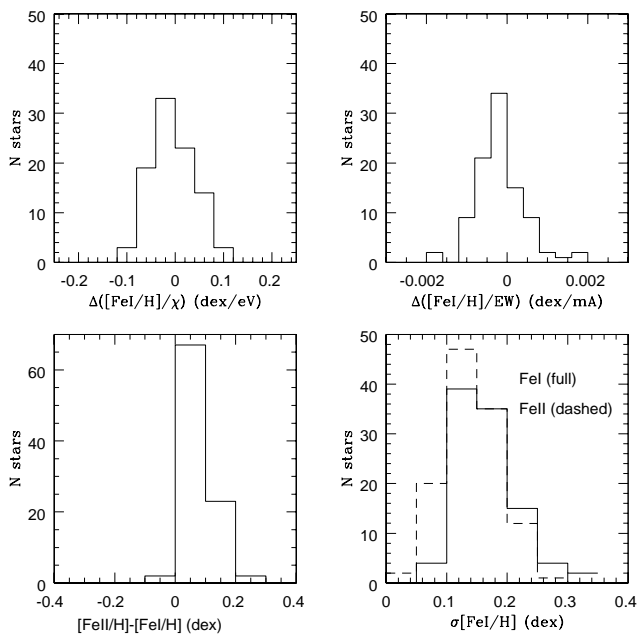}
\caption{Global quality of the stellar parameters over the sample. The upper two panels display the distribution of the slopes for [\ion{Fe}{I}/H] abundances with respect to the excitation potential $\chi_{ex}$ (left); and EWs of lines (right). The lower two panels show the distribution of the ionisation balance, $\rm [\ion{Fe}{I}/H]-[\ion{Fe}{II}/H]$, (left); and dispersion of Fe abundances from individual lines around the mean (right).}
\label{fig_diagnostics2} 
\end{figure}

\subsection{Stellar parameters - spectroscopy}

Iron lines, \ion{Fe}{I} and \ion{Fe}{II}, were identified (see Table~\ref{table:linelist}), measured in all spectra, and used to constrain the stellar parameters. Model atmospheres are OSMARCS models kindly provided by Plez (private communication
2000-2002), computed with the MARCS code, initially developed by \citet{Gustafsson75} and subsequently updated by
\citet{Plez:1992}, \citet{Edvardssonetal:1993} and \citet{Asplundetal:1997}. In the metallicity range $\text{[Fe/H]}<-1$, this grid assumes a standard  $\text{[$\alpha$/Fe]}=+0.4$, which overestimates the actual  [$\alpha$/Fe] in Sculptor for stars with metallicities $\text{[Fe/H]}>-2$. The metallicities assumed in the models were therefore corrected to account for this effect, by lowering the actual iron abundance of the star by a factor ensuring that the overall metallicity of the star was conserved\footnote{For a star of a given [Fe/H], $\rm Fe_{mod} = [Fe/H]-0.3([Fe/H]+2.0)$, i.e. at $\rm[Fe/H]=-1$, the model is assumed to be $-0.3$\,dex less than the actual Fe abundance of the star, and at $\rm [Fe/H]=-2$, $\rm Fe_{mod} = [Fe/H]$.}. 

Abundance calculations were performed using CALRAI, an LTE spectrum synthesis code originally developed by \citet{Spite67}, with numerous updates and improvements over the years. Abundances from individual lines are computed, and the measurement uncertainty on each EW ($\delta_{\rm DAO}$, estimated by DAOspec) is propagated into an uncertainty in the resulting abundance for each line. The error estimates on abundances are then carried throughout the stellar parameter and abundance derivation, by weighting each line by $1/(\delta_{\rm DAO})^2$ in the computations of slopes or means.

The curve of growth for Fe was examined for each star as a final check that both the \ion{Fe}{i} and \ion{Fe}{ii} lines are well fitted using the adopted parameters for all line strengths. 
Measurements of Fe as a function of $\lambda$, line excitation potential and EW for the adopted stellar parameters are shown for three typical stars in Fig.~\ref{fig_diagnostics1}.
An overview of the relevant tests for Fe measurments in GIRAFFE and UVES is shown in Fig.~\ref{fig_diagnostics2}, which includes the distribution of the slopes for [\ion{Fe}{i}/H] abundances with respect to the line excitation potential and EW; the distribution of residual [\ion{Fe}{i}/H]-[\ion{Fe}{ii}/H] (ionisation balance); and the distribution of observed dispersion of Fe abundances from individual lines around the mean ($\sigma_{\rm X}$). There is a slight shift in the distributions of  $\sigma_{\rm \ion{Fe}{i}}$ and $\sigma_{\rm \ion{Fe}{ii}}$, arising from the fewer lines of \ion{Fe}{II} measured. 

\subsubsection{Microturbulence velocities \vt} 
The microturbulence velocities, \vt, were determined by requiring a match between the \ion{Fe}{I} abundances and their expected EWs. Using expected EWs rather than the observed strength of the line removes a bias towards higher \vt\, which is created by the correlated errors between measured EWs and Fe abundances of individual lines, as first highlighted by \citet{Magainetal:1984} and more recently explored by \citet{Hill11}. This also allows more efficient identification of false detections of faint lines. The Fe abundance was then checked against that adopted for these initial calculations, then iterated until the model metallicity matched the final measured \ion{Fe}{i} abundances. The uncertainty on \vt, for each star, was evaluated from the uncertainty in the slope of the \ion{Fe}{I} abundances with line strength. The final \vt\ uncertainties are on average $\pm$0.20~km$\,$s$^{-1}$.

\subsubsection{Effective temperatures \teff} 
The initial photometric estimates of \teff\ were checked by examining the relation between the \ion{Fe}{i} line abundances and the excitation potential, $\chi$. The result was re-examined for any star with a slope $\ge 2\sigma$. This included 25 stars of the 89 GIRAFFE targets, and 1 of the UVES targets. In the majority of the cases, the slopes were found to be simply due to a large dispersion in the \ion{Fe}{i} abundances. For 11 GIRAFFE and 1 UVES targets however, the initial temperature estimates (from photometry) were adjusted to provide an acceptable excitation equilibrium. These adjustments were in random directions, and all within 100~K of the initial temperature except in two cases, ET0330 which required a $-150$~K temperature decrease and ET0241 a $+200$~K temperature rise. The latter, ET0241, only had available temperature from one colour, (V-I), while ET0330 had also the IR photometry.

\subsubsection{Surface gravities \logg} 
The photometric estimates of \logg\ were adjusted to ensure that the same abundance of iron is determined from the neutral and ionised Fe lines, within uncertainties. More precisely, we required that $|\rm[\ion{Fe}{I}/H]-[\ion{Fe}{II}/H]|\leq 2\times\rm \sqrt(\sigma_{\rm \ion{Fe}{I}}^2+\sigma_{\rm \ion{Fe}{II}}^2)$. These spectroscopic \logg\ values were adopted in the abundance analysis, and are listed in Table~\ref{table:Stellpar}. The uncertainty on \logg\ was evaluated from the uncertainties on the \ion{Fe}{I} and \ion{Fe}{II} abundances, and is on average 0.31\,dex. Our spectroscopic values have a lower limit, $\text{\logg}\ge0.0$, due to the limits of the available grid of stellar atmosphere models. Only six stars actually hit this limit, and of those, only two have Fe out of ionisation equilibrium (see Table~\ref{table:Stellpar}).

\section{Abundance determinations}

The FLAMES/GIRAFFE spectra present some challenges because of the limited wavelength coverage and rather low spectral resolution  \citep[e.g.][]{Pompeiaetal:2008, Letarte10}, compared to that used by classical HR abundance analysis. To ensure homogeneity, however, we have chosen to perform an analysis as similar as possible for our GIRAFFE and UVES spectra.

The abundances of the chemical elements were determined from EW measurements, which are listed in table~\ref{table:ew}. Hyperfine structure (HFS) corrections were included for: \ion{Ba}{II} 6141 and 6496~\AA\ (\citealt{Rutten:1978}, the isotopic solar composition from \citealt{McWilliam:1998}); \ion{La}{II} 6320~\AA\ (\citealt{Lawleretal:2001b}, with the oscillator strength from \citealt{Bordetal:1996}); and \ion{Eu}{II} 6645~\AA\ (\citealt{Lawleretal:2001a}, using the oscillator strength from \citealt{Shetrone01}). The HFS corrections are small or negligible for these lines, ranging from zero to 0.14~dex, with the strongest dependence on the line strength. HFS corrections were also computed for the \ion{Co}{I} 5483 line (using atomic data from \citealt{Prochaskaetal:2000}), which proved to be larger (ranging from 0.03 to 1.0 dex) and primarily dependent on both line strength and \vt. HFS was not included for the \ion{Ba}{II} line at 5854~\AA\, which was only available for the UVES spectra. The effects are expected to be small (e.g. \citealt{MashonkinaZhao2006}), and in fact it agreed with  the other lines, with no significant offset. 

The most metal-poor stars in our sample ($\rm[Fe/H] \le -2.2$) all happen to have been observed with GIRAFFE. The weak spectral lines coupled with the somewhat limited spectral resolution of the GIRAFFE spectra make these measurements less reliable. Thus, we have taken extra care to analyse these stars. We note that these stars have metallicities in agreement with the \ion{Ca}{II} triplet results from our LR survey \citep{Battaglia08b}. No corrections have been made to our abundances for non-LTE effects. We have attempted to compare our abundances with similar LTE analyses to minimise this source of error.

\begin{table}
\caption{\label{table:errormod} Abundance errors arising from uncertainties in stellar parameters over our full sample. The average uncertainties of our stellar parameters: $\delta \text{\teff}=\pm100$~K, $\delta\text{\logg}=\pm0.31$, $\delta\text{\vt}=\pm0.20$}.
\centering
\begin{tabular}{lccc}
\hline\hline
[X/Y] & $\Delta [X/Y]_\text{\teff,\logg}$ & $\Delta [X/Y]_\text{vt}$  & $\sigma_{\rm mod}$ \\
\hline
 [\ion{Fe}{I}/H]  &  +0.13   & $-0.08$ &    0.16  \\
 
 [\ion{O}{I}/Fe]  &  +0.06   &  +0.02 &    0.06  \\
 
 [\ion{Na}{I}/Fe] & $-0.06$   &  +0.08 &    0.10  \\

 [\ion{Mg}{I}/Fe]   & $-0.11$   & $-0.01$ &    0.11  \\

 [\ion{Al}{I}/Fe] & $-0.06$   &  +0.08 &    0.11  \\

 [\ion{SI}{I}/Fe] & $-0.10$   &  +0.08 &    0.13  \\

 [\ion{Ca}{I}/Fe] & $-0.03$   &  +0.02 &    0.05  \\

 [\ion{Sc}{II}/Fe]& $-0.02$   &  +0.07 &    0.08  \\

 [\ion{Ti}{I}/Fe] &  +0.04   &  +0.06 &    0.08  \\

 [\ion{Ti}{II}/Fe]& $-0.04$   &  +0.02 &    0.06  \\

 [\ion{Cr}{I}/Fe] &  +0.06   & $-0.01$ &    0.07  \\ 

[\ion{Co}{I}/Fe] &  +0.02   &  +0.02 &    0.05  \\

[\ion{Ni}{I}/Fe] & $-0.02$   &  +0.02 &    0.08  \\

[\ion{Zn}{I}/Fe] & $-0.12$   &  +0.03 &    0.13  \\

[\ion{Ba}{II}/Fe]&  +0.01   &$ -0.10$ &    0.11  \\

[\ion{La}{II}/Fe]&  +0.02   &  +0.07 &    0.08  \\

[\ion{Eu}{II}/Fe]& $-0.02$   &  +0.06 &    0.07  \\
\hline
\end{tabular}
\end{table}

\subsection{Error estimates}

Uncertainties on individual elemental abundances were estimated from three different sources: 
\begin{itemize}
\item Individual errors on EW measurements are given by DAOspec and are propagated through the abundance calculations to produce an individual error on each single line measurement  ($\rm \delta_{DAOi}$ for each line $i$), and propagated on the mean abundance for each element X, $\rm \delta_{DAO}(X)$.
\item The dispersion ($\rm \sigma_{obs}$) around the mean abundance of a given species measured from several lines reflects a combination of line measurement errors, uncertainties on atomic data and modelling errors.
\item Abundance errors due to uncertainties in the stellar parameters of the targets were estimated by re-computing abundances with varying stellar parameters (\teff, \logg, \vt), according to the individual error estimates on the stellar parameters. Because of the strong covariance between \teff\ and \logg, astrophysically bound by stellar evolution, we varied \teff\ and \logg\ in lock-step while \vt\ was varied on its own. The overall error due to stellar parameter uncertainties ($\rm \sigma_{mod}$) is then computed as the quadratic sum of the uncertainties arising from (\teff\ + \logg)  and \vt. Table~\ref{table:errormod} reports the mean over the sample of these errors.
\end{itemize}

\noindent The line measurement and atomic data uncertainties were combined into an observational error on the abundance of element~X: 
$$\rm err_{obs}(X) = max (\delta_{DAO}(X), \sigma_{obs} /\sqrt{N_X} )$$
where $\rm N_X$ is the number of lines measured for element X and $\rm \sigma_{obs}$ is set to $\rm \sigma_{obs}(X)$ if $\rm N_X > 3$  or to $\rm \sigma_{obs}(Fe)$ if $\rm N_X \leq 3$. That is, we use the dispersion of iron around the mean in each star as a surrogate for the dispersion around the mean abundance when too few lines of element X are available to robustly estimate this dispersion. This observational error $\rm err_{obs}(X)$ is then combined quadratically with the overall error due to stellar parameters $\rm \sigma_{mod}$ to estimate the final error on abundances, provided in Table~\ref{table:abund}.

\subsection{Verification of the abundance analysis}

Several tests were made to ensure that the abundance analysis of the FLAMES/GIRAFFE spectra was reliable. For this purpose, six stars with previously published analysis from UVES slit spectra \citep{Shetrone03,Geisler05} were reobserved with GIRAFFE. In these tests we compared: stellar abundances; EW measurements; and elemental abundance results, between present and previous work. In addition, we made a comparison with the results for the carbon-enhanced metal-poor (CEMP-no) star ET0097, obtained with UVES slit spectroscopy \citep{Skuladottir15b}. This verification process showed the results of our GIRAFFE analysis to be reliable. For more details see Appendix~A.

\section{Results}

Elemental abundances have been measured for 89 (82 new) stars in the Sculptor dSph from FLAMES/GIRAFFE and 10 new stars with FLAMES/UVES spectroscopy. We have focused our attention on seventeen elements to characterise the light elements (Li, Na), $\alpha$-elements (O, Mg, Si, Ca, Ti), iron-peak elements (Sc, Cr, Fe, Co, Ni, Zn), and heavy elements (Ba, La, Nd, Eu). The results of the abundance measurements are listed in Table~\ref{table:abund}.

\begin{figure}
\centering
\includegraphics[width=\hsize-1cm]{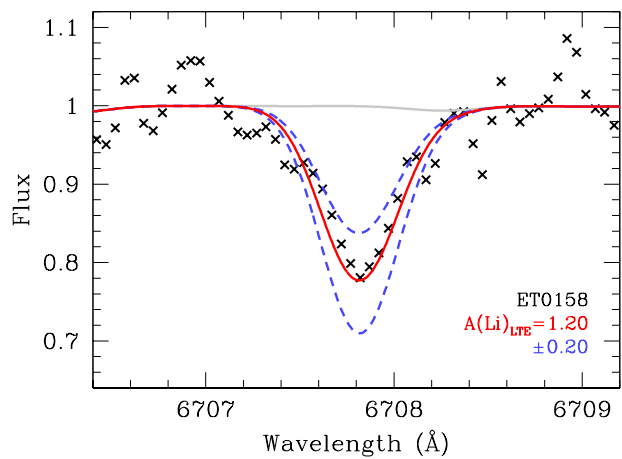}
\caption{Spectra of the \ion{Li}{I} line at 6707.8~\AA\ in the star ET0158. Black crosses show our GIRAFFE spectrum, the red line is the best fit at $\text{A(Li)}_\text{LTE}=1.20$, and blue dashed lines show $\pm0.2$~dex from this value. The grey solid line is the case with no Li present.
}
\label{fig:ET0158}
\end{figure}

\subsection{Li detection}

As Li is destroyed in stellar interiors, Li-poor material is mixed up to the surface at later stages of stellar evolution (e.g. \citealt{Gratton00}), and Li abundances in giant stars are typically very low. However, Li-enhanced giants have been found in moderate numbers in various environments (\citealt{Monaco08,Gonzalez09,Monaco11,Ruchti11,Kirby12,Kirby16,Martell13,Liu14,Casey16,Mena16}). Explaining the high Li in these stars requires either a mechanism to avoid depletion or an extra source of Li, apart from the amount the star was born with. 

In our FLAMES/GIRAFFE+UVES sample of 99 giant stars in Sculptor, we could detect Li in only one star, ET0158, see  Fig.~\ref{fig:ET0158}, which has $\text{A(Li)}_\text{LTE}=1.20\pm0.26$, and a metallicity of $\text{[Fe/H]}=-1.80\pm0.21$. Applying NLTE-corrections provided by \citet{Lind09a} results in $\text{A(Li)}_\text{NLTE}=1.40\pm0.26$ for ET0158.
The detection limit in our sample was $\leq0.5$~dex in the mean, so for this sample of giant stars in Sculptor with $V\lesssim18.4$ (or $M_V\lesssim-1.1.$), we estimate a fraction of $1\%^{+2.3}_{-0.8}$ (errors from \citealt{Gehrels86}) of the stars to have $\text{A(Li)}_\text{LTE}>0.5$.

\subsubsection{Comparison with literature data}

\begin{figure}
\centering
\includegraphics[width=\hsize-0.5cm]{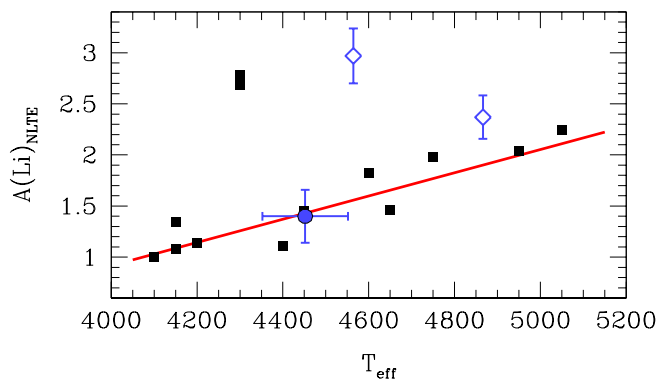}
\caption{Lithium detections in Sculptor giants (blue) as a function of \teff. The filled circle is ET0158, and open diamonds are two Li-rich giants from \citet{Kirby12}. The red line shows the trend found by \citet{Gonzalez09} for Galactic bulge giants (black squares) with $1\leq\text{A(Li)}_\text{NLTE}\leq2.5$.
}
\label{fig:TLi}
\end{figure}

In a sample of $\approx$400 giant stars in the Milky Way bulge, \citet{Gonzalez09} found 13 Li-detections ($\approx$3\%). Two of these stars have very high values,  $\text{A(Li)}_\text{NLTE}>2.5$, but for the other 11 stars a correlation between \teff\ and Li abundance was found (also seen for different samples in \citealt{Brown89,Pilachowski90,Pilachowski00,Lebzelter12}). Somewhat surprisingly, the Sculptor star ET0158, seems to fall directly onto this relation, see Fig.~\ref{fig:TLi}. Compared to the bulge sample, ET0158 has very different metallicity and luminosity, so this is not necessarily expected. With a sample of only one star it is quite possible, however, that ET0158 lands on this relation by mere chance. Also included in Fig.~\ref{fig:TLi} are two Li-rich giants in Sculptor from Keck~II DEIMOS medium-resolution spectroscopy  \citep{Kirby12}. The more Li-rich of these two clearly falls off the relation, in a similar way to the two Li-rich giants in the bulge sample, while the other one is more ambiguous.

\begin{figure}
\centering
\includegraphics[width=\hsize-0.5cm]{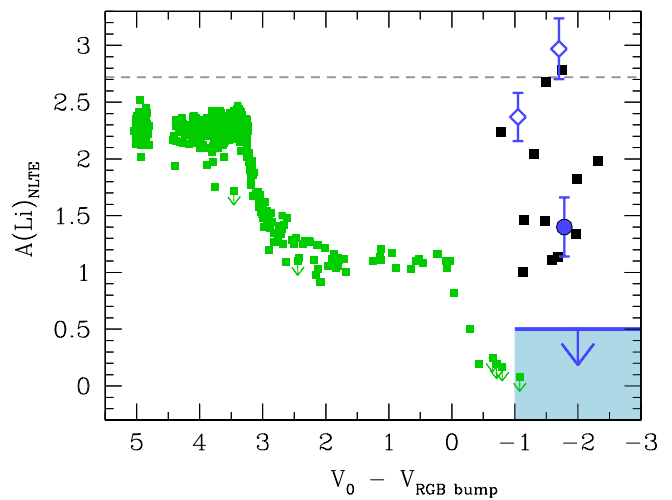}
\caption{Lithium abundances as a function of ${V_0}$ relative to the predicted $V_\text{RGB bump}$ \citep{Ferraro99}, for Sculptor and Milky Way bulge giants (same symbols as in Fig.~\ref{fig:TLi}). The typical detection limit of our FLAMES sample, $\text{A(Li)}_\text{LTE}=0.5$, is shown with a blue line, and upper limits for non-detections lie in the shaded region. The globular cluster NGC~6397 is also shown with small green squares \citep{Lind09b}. The primordial Li abundance is shown with a grey dashed line \citep{Coc12}. 
}
\label{fig:VLi}
\end{figure}

Following the approach of \citet{Kirby12}, the Li-abundance is plotted in Fig.~\ref{fig:VLi}, as a function of the de-redenned magnitude of the star, relative to the RGB bump luminosity ($V_{\rm o}-V_\text{RGB bump}$). The predicted $V_\text{RGB bump}$ is calculated for the Sculptor and bulge stars according to \citet{Ferraro99}, using the stellar metallicities and assuming an average age of 10~Gyr. This choice of age is justified by the fact that the Milky Way bulge is predominantly old (e.g. \citealt{zoccali03,Bensby17,Bernard18}), and Sculptor is also dominated by an old population (e.g. \citealt{deBoer12}). A change in $\pm2$~Gyr gives a shift in $V_\text{RGB bump}$ of $\pm0.1$~dex. The reddening towards the bulge is adopted from \citet{zoccali03} and for  Sculptor, is the same as listed in Section~\ref{sec_photAPs}.

Although similar trends to that in Fig.~\ref{fig:TLi} have been found in different stellar samples \citep{Brown89,Pilachowski90,Pilachowski00,Lebzelter12}, it is generally offset to that found in \citet{Gonzalez09}. As the $V_\text{RGB bump}$ is very metallicity dependent, the Sculptor and bulge samples overlap in $V_0-V_\text{RGB bump}$, despite the different intrinsic luminosities. Combined with the similar expected ages, and thus similar masses, this seems to indicate that these samples 
catch the giants stars in a similar phase of their internal mixing history (as traced by their luminosity above the RGB bump),
potentially explaining the relation found in Fig.~\ref{fig:TLi}, although this needs to be confirmed with a larger sample of measurements in Sculptor.

\subsubsection{Possible explanations}

Many different scenarios have been invoked to explain the unexpectedly high Li-abundances observed in a small fraction of giant stars. Here we will discuss those scenarios having observable consequences which can be checked in the data for this particular star, ET0158:

\begin{itemize}
\item \textit{Binary companion:} Giant stars in binary systems have been shown to have Li abundance to \teff\ relation, similar to that shown in Fig.~\ref{fig:TLi}; and for close binaries Li depletion seems uncommon \citep{Costa02}. In four velocity measurements, from spectra taken in 2003-2013, ET0158 shows no evidence of being in a binary system \citep{Skuladottir17}. With the limited data a binary companion cannot be excluded, but we note that in \citet{Gonzalez09}, only one star showed significant velocity variations, so the scenario where all of their sample stars were in a binary system is not favoured. \\

\item \textit{Mass loss:} High Li abundances have been linked to the evolution of circumstellar shells \citep{Reza96,Reza97}. Within this scenario, an infrared excess is expected, as well as asymmetries in the H$_{\alpha}$ profile, neither of which is observed in ET0158. However, recent studies also seem to indicate that high Li abundances and infrared excess are not necessarily correlated \citep{Kumar15}.\\

\item \textit{Rapid rotator:} When infrared excess and asymmetric H$_{\alpha}$ profile are present, there is a clear relation between high rotational velocities and very high Li abundances for K giant stars \citep{Drake02}. ET0158 shows no signature of being rapidly rotating, as its FWHM is within (and even slightly below) what is normal for the Sculptor sample.\\

\item \textit{AGB star:} Asymptotic giant branch (AGB) stars can generate Li (e.g. \citealt{Cameron71,Cantiello10}), so if ET0158 is an early AGB star, that could explain the measured Li abundance. This theory is supported by the star's colour, which is slightly bluer than typical for the sample, see Fig.~\ref{fig:cmd}. This results in a relatively young age estimate, $7.6\pm 1.6$~Gyr, for a star of this metallicity in Sculptor: $<\text{Age}>~=~9.7\pm0.5$ for stars where [Fe/H] is within $\pm0.2$~dex from that of ET0158. In support of this explanation, \citet{Kirby16} found a higher fraction of Li-enhancement among AGB stars ($1.6\pm1.1\%$) compared to RGB stars ($0.2\pm0.1\%$) in their survey of 25 globular clusters.
\end{itemize}

\noindent For a more detailed discussion of the suggested mechanisms for Li-enhancement in giant stars, we refer to \citet{Gonzalez09,Kirby16,Aguilera-Gomez16}, \citet{Fu18}, and \citet{Bensby18}.

\begin{figure*}
\centering
\includegraphics[width=\hsize-2.5cm]{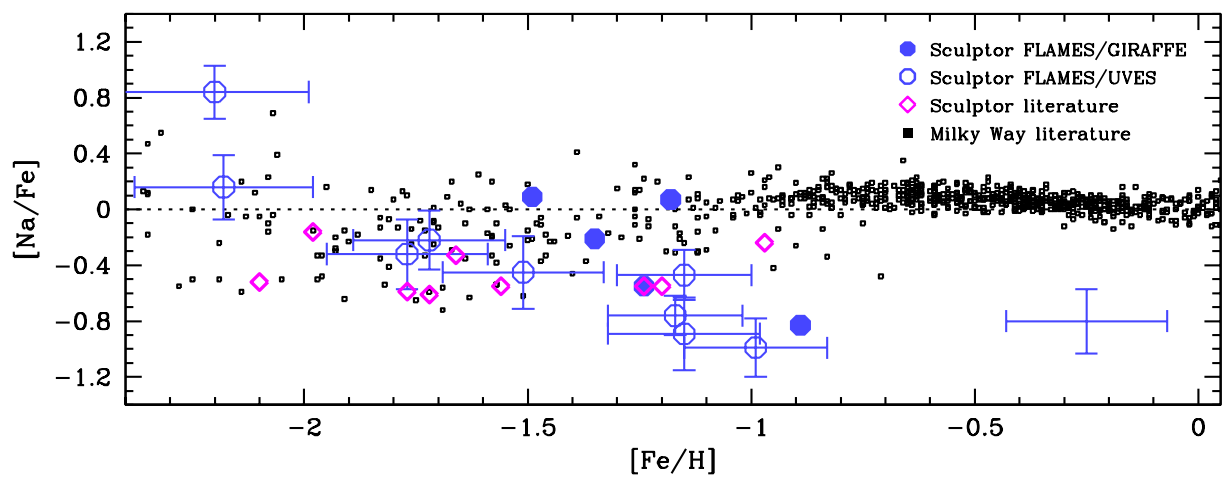}
\caption{ [Na/Fe] as a function of [Fe/H] for stars in Sculptor and the Milky Way. The blue circles are Sculptor stars from this work, GIRAFFE (filled) and UVES (open). Representative error bars for the GIRAFFE data is shown in blue (bottom right corner). Magenta open diamonds are previously published Sculptor stars, and the Milky Way is shown with small black squares. References: \textit{Sculptor:} \citealt{Shetrone03,Geisler05,Frebel10,Starkenburg13,Skuladottir15b,Simon15,Jablonka15}. \textit{Milky Way:} \citealt{Venn04} compilation; \citealt{NissenSchuster10}.
}
\label{fig:odd}
\end{figure*}

\begin{figure*}
\centering
\includegraphics[width=\hsize-1cm]{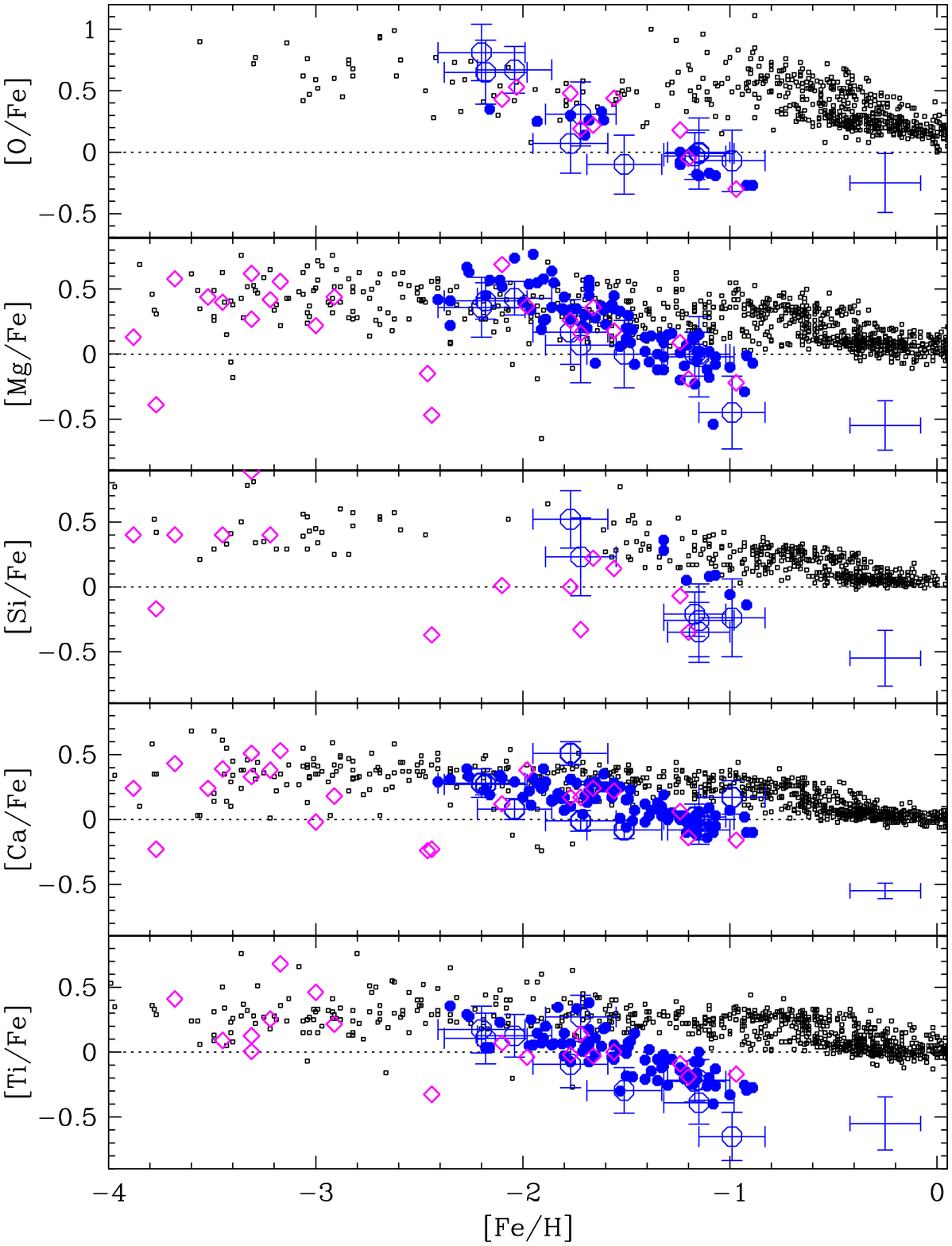}
\caption{Abundance ratios for $\alpha$-elements as a function of [Fe/H] for Sculptor and the Milky Way. Symbols and Sculptor references are the same as in Fig.~\ref{fig:odd}. \textit{Milky Way:} \citealt{Fulbright00} (Si); \citealt{Carretta00} (O); \citealt{Nissen02} (O); \citealt{Reddy03,Reddy06} (Si); \citealt{Cayrel04} (O,~Mg, Si, Ca, Ti); \citealt{Venn04} compilation (Mg, Ca, Ti);  \citealt{Bensby05} (O, Si); \citealt{GarciaPerez06} (O); \citealt{Ramirez07} (O); \citealt{NissenSchuster10} (O, Si, Ca, Ti). Only O abundances derived from the [\ion{O}{I}] line were included.
}
\label{fig:alpha}
\end{figure*}

\begin{figure*}
\centering
\includegraphics[width=\hsize-1cm]{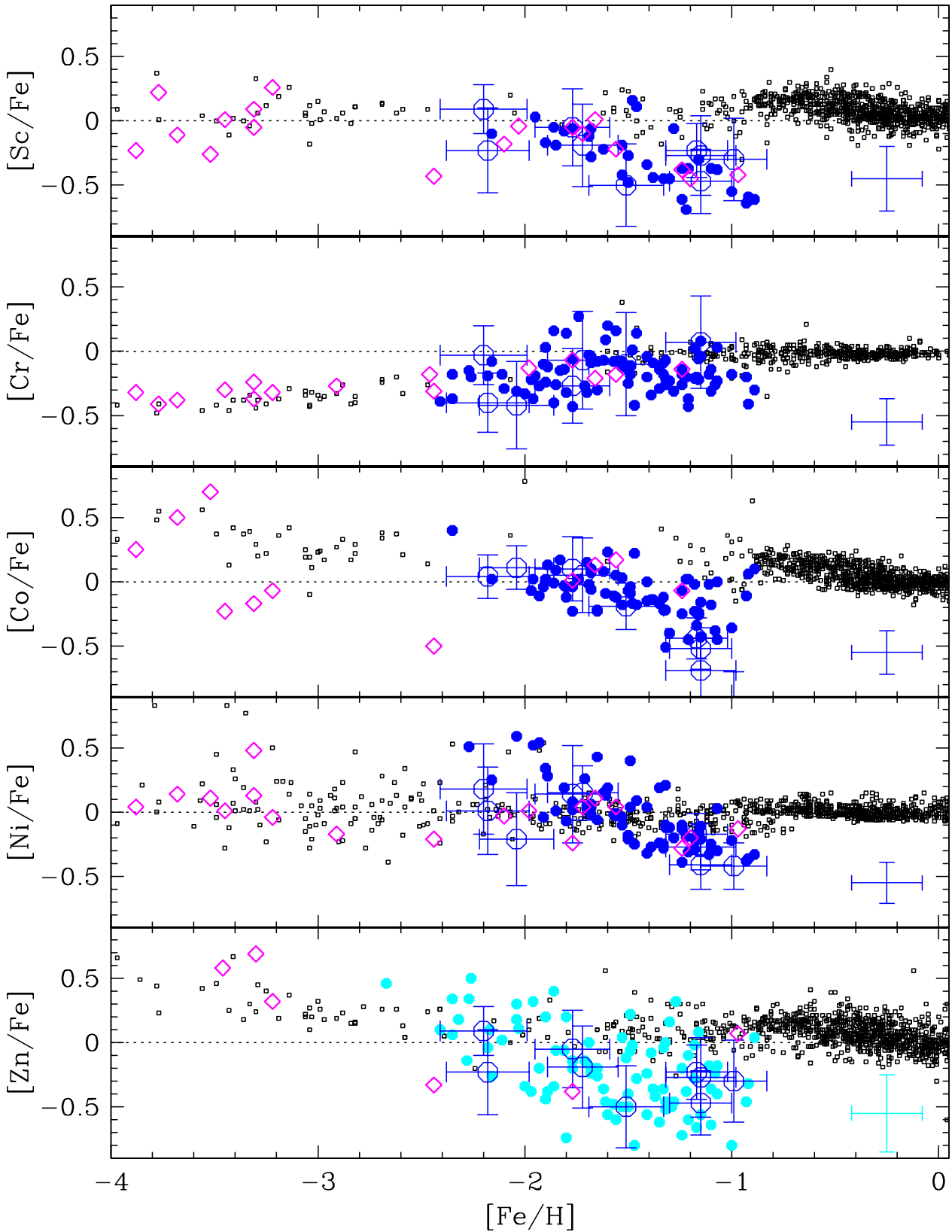}
\caption{Abundance ratios for iron-peak elements as a function of [Fe/H] for Sculptor and the Milky Way. Cyan filled circles at the bottom panel are Zn measurements for Sculptor from \citet{Skuladottir17} which include the stars in our GIRAFFE sample, and a representative error bar is also shown in cyan. Otherwise, symbols and Sculptor references are the same as in Fig.~\ref{fig:odd}. \textit{Milky Way:} \citealt{Fulbright00} (Cr, Ni); \citealt{Reddy03,Reddy06} (Sc, Cr, Co, Zn); \citealt{Venn04} compilation (Ni); \citealt{Cayrel04} (Cr, Co, Ni, Zn); \citealt{NissenSchuster10,NissenSchuster11} (Cr, Ni, Zn); \citealt{Ishigaki13} (Zn), \citealt{Bensby14} (Zn), \citealt{Barbuy15} (Zn).
}
\label{fig:iron}
\end{figure*}

\begin{figure*}
\centering
\includegraphics[width=\hsize-1cm]{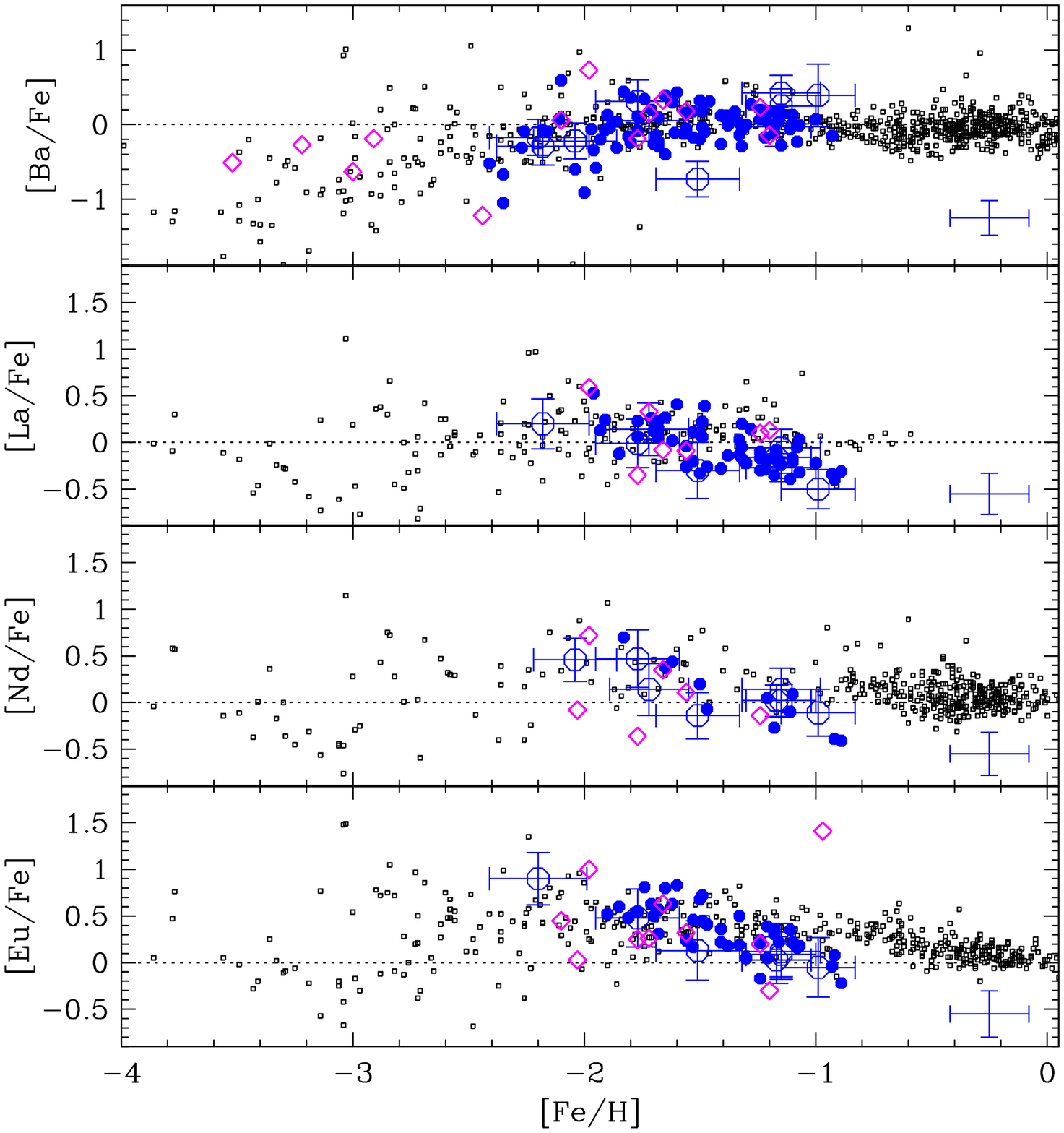}
\caption{Abundance ratios for heavy elements as a function of [Fe/H] for Sculptor and the Milky Way. Symbols and Sculptor references are the same as in Fig.~\ref{fig:odd}. \textit{Milky Way:} \citealt{Burris00} (Ba, La, Nd, Eu); \citealt{Reddy03,Reddy06} (Nd); \citealt{Venn04} compilation (Ba, La, Eu); \citealt{Simmerer04} (La, Eu); \citealt{Francois07} (Ba, La, Nd, Eu).\\
}
\label{fig:heavy}
\end{figure*}

\subsection{The odd elements Na and Al}

The only  \ion{Na}{I} lines accessible in the GIRAFFE spectral range are the \ion{Na}{i} doublet at 6154 and 6161~\AA. In our sample these lines are very weak and only detectable in a few stars, as shown in Fig.~\ref{fig:odd}. With the larger wavelength coverage of UVES, more lines were accessible, see Table~\ref{table:linelist}. The [Na/Fe] abundance ratios seem to be slightly higher in the GIRAFFE sample compared to UVES, however, no systematic difference was found in abundance analysis from different Na lines in the UVES spectra. One possible reason for this offset could be that the lines are close to the detection limit in the GIRAFFE spectra, so only detected when the Na abundance tends to be high.

Two \ion{Al}{I} lines, at 6696 and 6699~\AA, were covered both by the UVES wavelength range and the HR15 setting in GIRAFFE. These very weak lines were only reliably detected in one GIRAFFE spectrum, ET0137, the most metal-rich star in our sample, with $\text{[Al/Fe]}=-0.35\pm0.27$, and in none of the UVES spectra.

\subsection{The $\alpha$-elements}
The O, Mg, Si, Ca and Ti abundances, are shown in Fig.~\ref{fig:alpha}. With the exception of Si, GIRAFFE and UVES measurements are in very good agreement. In the case of Si, the GIRAFFE results are systematically shifted to higher abundance. This is the consequence of the line list, as only one \ion{Si}{I} line, at 6245~\AA, is accessible with the GIRAFFE spectra, while the line most commonly used for the UVES spectra is at 5685~\AA. In the UVES star ET0143 both of these lines were measured, but the redder one gave a result $+0.3$~dex higher compared to the one at 5685~\AA, thus explaining this difference. In the case of O, Mg, Si, Ca, Ti, the scatter was tested and found compatible with measurement uncertainties.

\subsection{Iron-peak elements}

Abundance ratios of the iron-peak elements Sc, Cr, Co, Ni, and Zn to Fe are shown in Fig.~\ref{fig:iron}. In all cases, GIRAFFE and UVES results are in very good agreement. The odd element Sc was measured using one relatively weak line at 6310~\AA, and could thus only be measured for high S/N GIRAFFE spectra, and typically not at the lowest metallicities. The heaviest of the iron-peak, Zn, was measured with a line at 4810 \AA\ in the UVES spectra. No Zn line was available with the GIRAFFE wavelength coverage of this work. However, Zn was measured from GIRAFFE spectra for $\approx$100 stars (85 overlapping with our sample) in \citet{Skuladottir17}, see more detailed discussion therein. The scatter in the iron-peak elements was found to be compatible with measurement uncertainties. However, there is a statistically significant correlation between the offsets from the mean trends in Ni and Zn, see further discussion in \citet{Skuladottir17}.

\subsection{Heavy elements}
Four heavy elements were measured, Ba, La, Nd and Eu, see Fig.~\ref{fig:heavy}. The GIRAFFE and UVES results are in good agreement for all four elements. Unlike the iron-peak and $\alpha$-elements, the scatter exceeds what is expected from measurement uncertainties for Ba. The lighter $n$-capture element Y was measured in the UVES sample and for four stars in the GIRAFFE samples, but this will be published with more Y measurements from complementary observations in the GIRAFFE HR7A setting in Sk\'{u}lad\'{o}ttir et al. (in prep.). Comparison of our Y measurement in ET0097 with that of \citet{Skuladottir15b} is done in Appendix A along with other elements for this star.

\subsubsection{Comparison with intermediate resolution spectroscopy}

A large number of stars (376) in the central field of Sculptor has previously been observed using Keck DEIMOS intermediate resolution spectra \citep{Kirby09, Kirby11}. Overall, their results show similar trends to those presented here. However, there are also some significant discrepancies. A larger scatter in abundance ratios is observed in the the Keck DEIMOS data (as expected from spectra of lower resolution and S/N), but there are also differences in trends, especially at the lowest metallicities. When all the \citet{Kirby09,Kirby11} data is considered, no knee  in the [$\alpha$/Fe] abundance ratios is observed, however, it does become visible when only their most reliable measurements are used. For a more detailed discussion of this, see Appendix~B. 


\begin{figure*}
\centering
\includegraphics[width=\hsize-2.5cm]{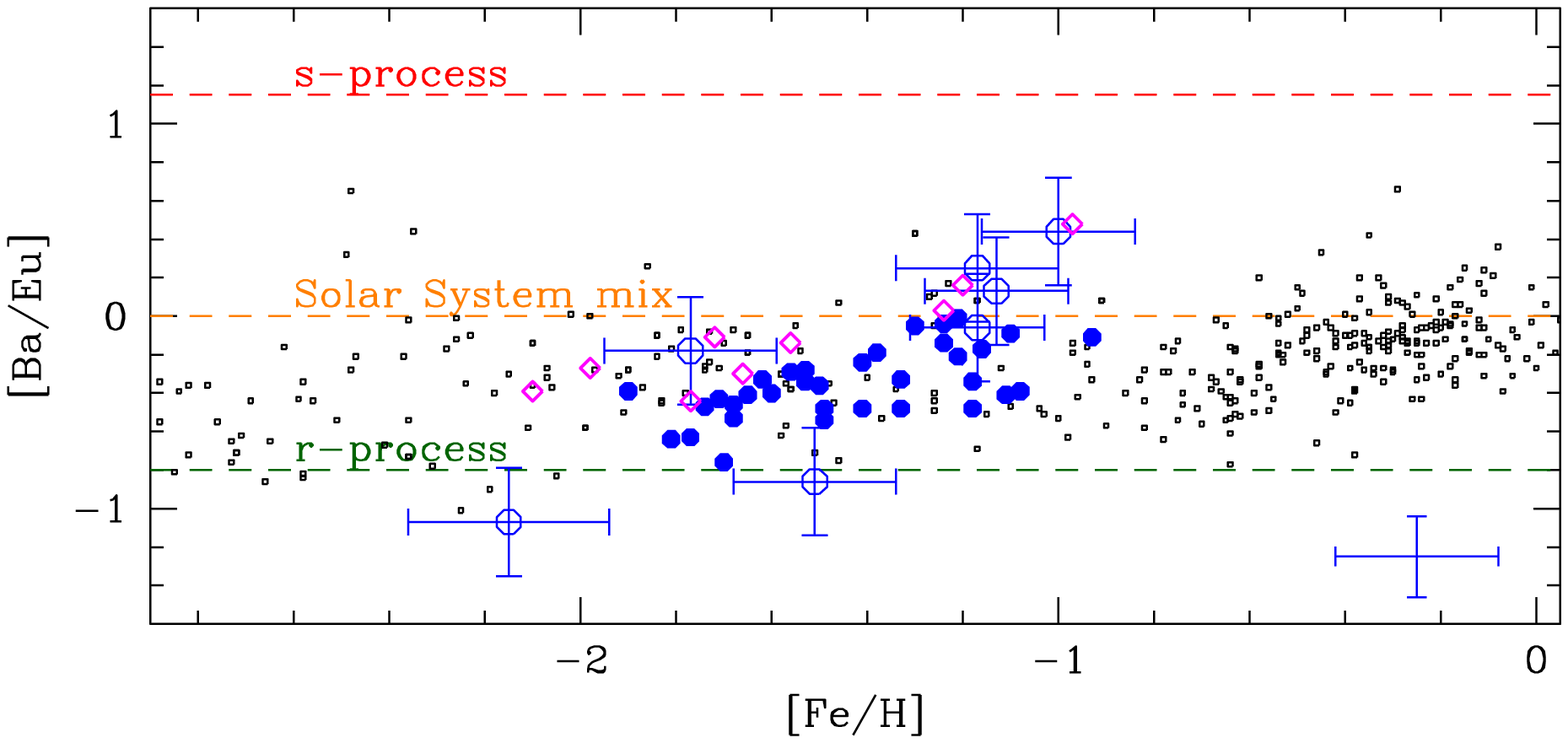}
\caption{Abundance ratios of Ba to Eu with [Fe/H], as tracers of the $s$- to $r$-process contributions to the heavy elements in Sculptor. The dashed lines show the solar [Ba/Eu] (orange), as well as the pure $r$-process (green) and pure $s$-process (red), from \citet{Bisterzo14}. Symbols and Sculptor references are the same as in Fig.~\ref{fig:odd}. \textit{Milky Way:} \citealt{Burris00}; \citealt{Venn04} compilation; \citealt{Francois07}. 
}
\label{fig:baeu}
\end{figure*}

\section{Sculptor, a textbook galaxy}

In many ways, the Sculptor dSph can be thought of as the ideal galaxy to study chemical evolution. It is small enough not to have the complicated structure of the Milky Way (bulge, thin/thick disks, halo), but large enough so that statistically significant samples of stars can be observed with HR spectroscopy, as presented here. It is a well studied galaxy, with a relatively simple star formation history, with a peak of star formation at the earliest times, which then steadily decreased until $\sim6$~Gyr ago \citep{deBoer12}, when star formation stopped. Thus its entire stellar population is old, dominated by stars of ages $>10$~Gyr. Sculptor can therefore be seen as a `textbook' galaxy, the ideal system to empirically witness chemical evolution reveal itself, from the earliest times to well after SN type~Ia and intermediate-mass stars started contributing the metal enrichment. 

\subsection{Abundance trends in Sculptor}

The chemical abundance ratios [X/Fe] as a function of [Fe/H] in Sculptor are significantly different from those observed in the Milky Way, see Figs.~\ref{fig:odd}, \ref{fig:alpha}, \ref{fig:iron}, and \ref{fig:heavy}. This suggests differences in the chemical enrichment histories of these two galaxies.

\subsubsection{General abundance trends}

Both in Sculptor and the Milky Way, supersolar values of $\text{$[\alpha$/Fe]}>0$ are observed at the lowest metallicities. This is consistent with initial pollution only from SNe type II, which explode on short timescales, $\approx10^{6}-10^{7}$~yr, and create large quantities of $\alpha$-elements, $\text{[$\alpha$/Fe]}>0$ (e.g. \citealt{Nomoto13}). After 1-2~Gyr, SN type Ia start to significantly contribute to the chemical evolution of each system, releasing primarily Fe and other iron-peak elements (e.g. \citealt{Iwamoto99}). This results in a knee in the [$\alpha$/Fe] abundance ratios, which start to decrease as the bulk of SNe type Ia start to contribute. In the Milky Way, this happens at relatively high metallicities, $\text{[Fe/H]}>-1$, but as Sculptor is a much smaller galaxy, with less efficient star formation, the gas is only enriched until $\text{[Fe/H]}\approx-1.8$ before the knee is observed, and the [$\alpha$/Fe] ratios start to decrease. Furthermore, the evolution and state of the gas in the galaxy will also affect how efficiently new metals are recycled into stars, ad might thus also influence the position of the knee \citep[e.g.][]{Lanfranchi07,Vinc16,Cote17,RomanoStarkenburg13,RevazJablonka12}.

The subsolar ratios of $\text{[$\alpha$/Fe]}$, seen at the highest metallicities in Sculptor, are typically not observed in the Milky Way disks or halo\footnote{with the exception of [O/Fe] at supersolar [Fe/H] in the Galactic disk \citep{Bensby14}.}, see Fig.~\ref{fig:alpha}. As star formation declined in Sculptor, the frequency of SN type~II gradually decreased. Due to the delayed timescales of SN type~Ia, however, their frequency at each time step is set by the higher star formation rate earlier on (typically 1-2~Gyr before). This could explain why the ratio of SN type~Ia to type~II in the later stages of the chemical evolution of Sculptor is relatively high \cite[e.g.][]{LanMatt03}. In the Milky Way disk, on the other hand, the contribution from SN type~Ia has always been together with a continuous contribution of SN type~II, and therefore the observed [$\alpha$/Fe] ratios are not as low. The slope of [$\alpha$/Fe] with [Fe/H] is therefore also steeper in Sculptor compared to the Milky Way, showing a very clear and unobscured signal of an increasing SN type~Ia contribution.

A similar declining trend can also be seen in Na and some of the iron-peak elements: Sc, Ni, Co and Zn (see Figs.~\ref{fig:odd} and \ref{fig:iron}). This indicates that the fraction of these elements to iron, [X/Fe], is higher in SN type~II than in SN type Ia at these metallicities in Sculptor. In the case of Na, some production from AGB stars is also expected \citep[e.g.][]{Karakas14}. However, considering the strong NLTE effects for Na lines in metal-poor giants (up to $\gtrsim0.5$~dex; e.g.~\citealt{Andrievsky07}), we advice against drawing strong conclusions for our limited number of LTE measurements of Na in Sculptor (see Fig.~\ref{fig:odd}).

\begin{figure}
\centering
\includegraphics[width=\hsize-0cm,clip=]{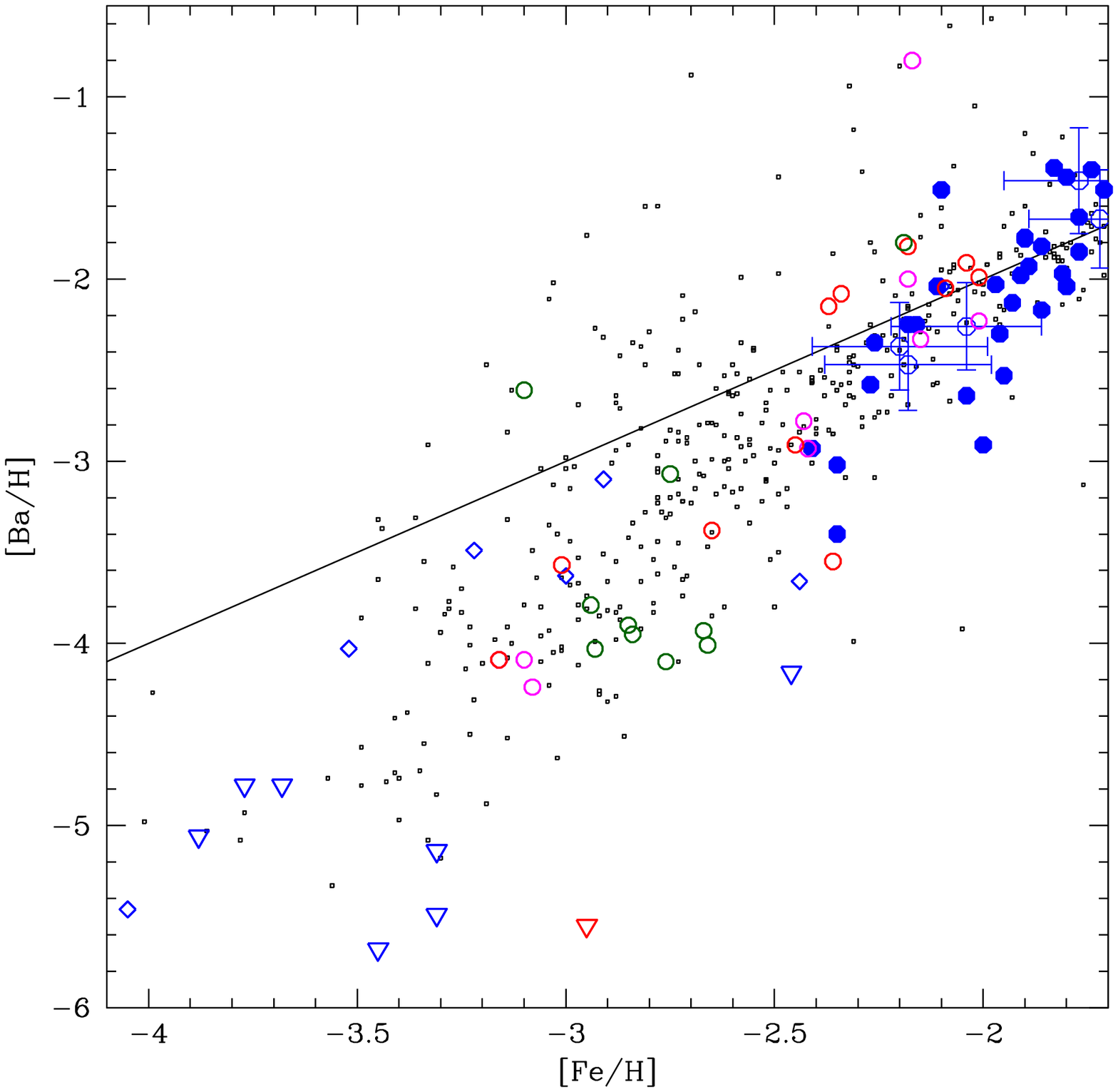}
\caption{[Ba/H] as a function of [Fe/H] in dSph galaxies. Sculptor is depicted with blue circles: GIRAFFE (filled); UVES (open with error bars). Literature samples for $\text{[Fe/H]}\leq-2$: Sculptor in blue open circles \citep{Shetrone03,Geisler05,Frebel10,Starkenburg13,Simon15,Jablonka15}; Draco in red \citep{Shetrone01,Cohen09,Tsujimoto17}; Sextans in green \citep{Tafelmeyer10,Aokietal:2009}; and Ursa Minor in magenta \citep{Shetrone01,Cohen10}. Inverted triangles indicate upper limits. Black points are Milky Way halo stars (\citealt{Burris00}; \citealt{Venn04}, compilation; \citealt{Barklemetal:2005}; \citealt{Francois07}). The line is $\text{[Ba/H]}=\text{[Fe/H]}$.
}
\label{fig:bah}
\end{figure}

\subsubsection{Abundance trends of the heavy neutron-capture elements}\label{sec:rs_trends}

The abundances of the heavy elements Ba, La, Nd, and Eu with [Fe/H] are shown in Fig.~\ref{fig:heavy}. These elements are created in the main rapid ($r$) and slow ($s$) neutron-capture processes. 

The heavy element Eu is mainly formed in the $r$-process, which produces more than 94\% of the Eu in the Sun \citep{Bisterzo14}. The $r$-process requires high-energy, neutron-rich environments \citep[e.g.][]{Sneden08} typically associated with the late evolution of massive stars, such as neutron star mergers (e.g. \citealt{Rosswog99,Wanajo14,Ishimaru15}); high energy winds accompanying core-collapse SNe (\citealt{Woosleyetal1994, QW03, Wanajo01, Wanajo13}); or magneto-hydrodynamical explosions of fast rotating stars (\citealt{Winteler12}). As measurements become challenging at the lowest metallicities, not much can be said about [Eu/Fe] at $\text{[Fe/H]}<-2$ in Sculptor. At higher metallicities, there is a decreasing trend of [Eu/Fe] with [Fe/H], similar to that of [$\alpha$/Fe]. This indicates that the $r$-process at these times and metallicities, was not sufficient to counteract the added contribution to Fe from SN type~Ia.

Conversely, the $s$-process dominates the production of Ba in the solar system, \citep[85\%,][]{Bisterzo14}. The $s$-process occurs in low mass ($M\lesssim4$~M$_\odot$) AGB stars \citep{Travaglio04}, and thus enters the evolution with a delay of at least $\sim$1~Gyr after the onset of star formation. At the earliest times in the Milky Way halo, the production of Ba is therefore dominated by the $r$-process. Early in the chemical evolution of Sculptor, $\text{[Fe/H]}\lesssim-1.8$, the [Ba/Fe] abundance ratios show a very large scatter, exceeding measurement uncertainties, with a subsolar mean value. Around $\text{[Fe/H]}\gtrsim-1.8$, 
the scatter in [Ba/Fe] decreases, and a plateau is reached around the solar value, in spite of the added Fe from SN type Ia at the same metallicities.

The relative $s$- to $r$-process contributions to the chemical enrichment of neutron-capture elements can be traced by [Ba/Eu], as is shown in Fig.~\ref{fig:baeu}. At the lowest metallicities in Sculptor, [Ba/Eu] is consistent with the $r$-process being the dominant production site of the neutron-capture elements. But as AGB stars start to contribute, the $s$-process gradually becomes more important, until at the highest metallicities solar or even supersolar ratios of [Ba/Eu] are reached. A similar trend appears at a higher metallicity in the Milky Way disk compared to Sculptor (analogous to the knee in [$\alpha$/Fe]). In this context, the rise of [Ba/Fe] in Sculptor (see Fig.~\ref{fig:heavy}) is clearly associated with the onset of the $s$-process. This rise in [Ba/Fe] happens at slightly higher metallicities in Sculptor than in the Milky Way halo. This was noted already by \citet{Tolstoy09}, and is a feature shared with other dSph galaxies \citep[e.g.][]{Shetrone01,Shetrone03}, although Sculptor is currently the galaxy that best samples the relevant metallicity regime ($\text{[Fe/H]}<-2$), together with Draco \citep{Tsujimoto17}. 

The other two elements in Fig.~\ref{fig:heavy}, La and Nd, are more evenly created by the $s$- and $r$-processes \citep[75\%  and 58\% of the solar La and Nd, respectively, come from the $s$-process, according to][]{Bisterzo14}. In the metallicity regime where the weak La and Nd lines could be measured ($\text{[Fe/H]}>-2$), the results indicate a slowly decreasing trend of [La/Fe] and [Nd/Fe].  This can be understood as being caused by the added SN type~Ia contribution to Fe in this metallicity range, partially compensated by the $s$-process.

The recent detection of the neutron-neutron star merger, GW170817 by the LIGO team \citep{Abbott17} and its ultraviolet, optical and infrared emission confirm neutron star (NS-NS) mergers as significant production sites for the $r$-process \citep{Chornock17,Cowperthwaite17,Drout17, Pian17,VIllar17}. But the question still remains, whether other proposed $r$-process sites also play a significant role. The dSph galaxies may be the best environment to figure out the dominant source(s) for their production. A wide range of works have examined the possibility that the enrichment of mini-halos by neutron star mergers are responsible for the large [$r$/Fe] dispersion in the Milky Way halo. The rare neutron star merger going off in a mini-halo would pollute it entirely, to a high level (e.g. \citealt{Ji16Nat, Ji16ApJ}) while others would hardly see any \citep[e.g.][]{Tsujimoto14, Hirai17b}.

 \citet{Tsujimoto17} examined the absolute [Eu/H] content of stars in the Draco dSph galaxy, and found them to align on two distinct plateaus, one high and one low, irrespective of their metallicity. They interpreted this as two separate discrete events, one that elevated the $r$-process elements to the level of the first plateau by producing a modest amount of $r$-process (which they associate to magneto-hydrodynamical explosion of a fast rotating star), and the second (a neutron star merger) which produced a much larger mass of $r$- process elements that were well distributed throughout Draco, elevating the level to the second plateau. 

In Fig.~\ref{fig:bah}, a sample of classical dSph galaxies show the run of [Ba/H] at $\text{[Fe/H]}<-2$, where the production of Ba is dominated by the $r$-process. When comparing Draco with other similar galaxies, it is not clear any more that this plateau-like behaviour of the $r$-process is a good description. In Sculptor, the [Ba/H] increase appears regular and does not follow steps. In Sextans and Ursa Minor, there are also no clear signs of a plateau either. Possibly these dSph galaxies are too large to suffer an extreme global enrichment as an UFD or a mini-halo might.

\begin{figure*}[htbp]
\centering
\includegraphics[width=\hsize-1.5cm,clip=]{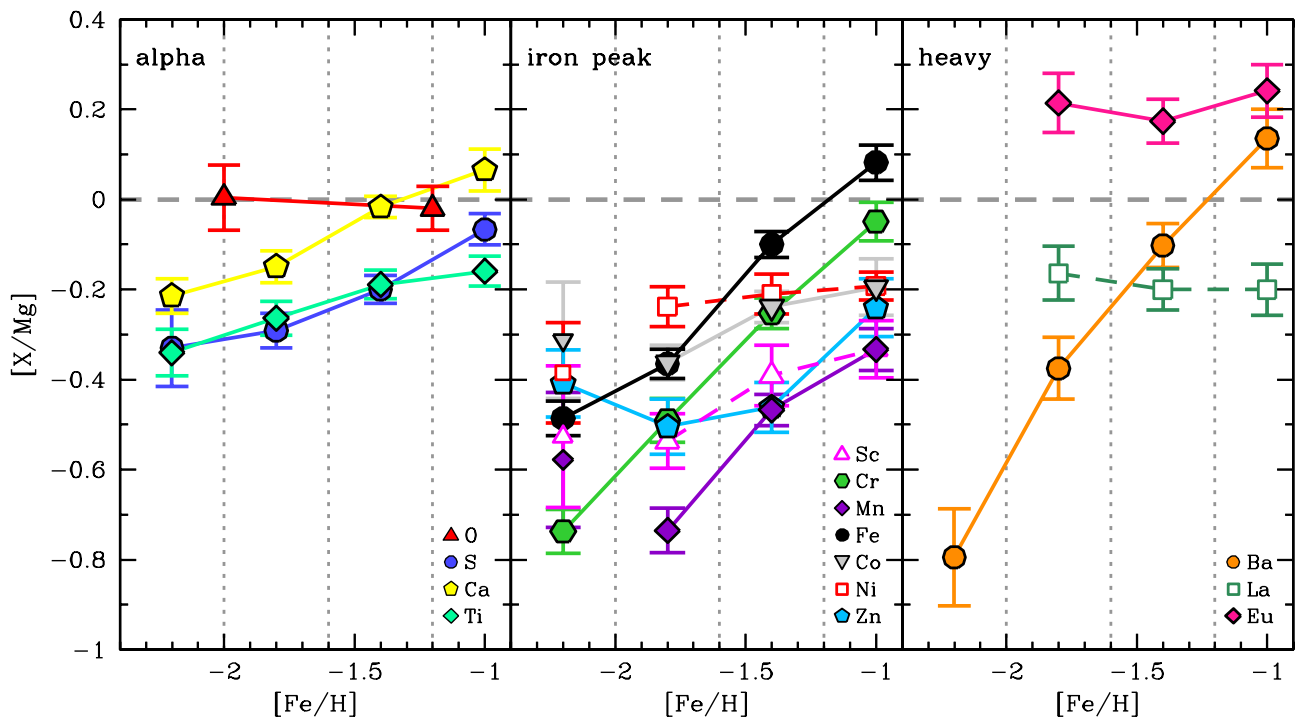}
\caption{Abundance ratios [X/Mg] as a function of [Fe/H], in three panels for: O, S$_\text{NLTE}$, Ca, Ti (left); Sc, Cr, Mn, Fe, Co, Ni, Zn (middle); and Ba, La, Eu (right). Data points including <10 stars (in all cases at the lowest metallicity bin) are not connected and are shown with small symbols for (number of stars): Sc (3), Mn (3), Co (4), and Ni (6). Elements only measured from very weak lines are shown with unfilled symbols (Sc, Ni, La). Error of the mean ($\sigma$/$\sqrt{N-1}$) is shown for each element with y-error bars. Dotted lines show the range of the [Fe/H] bins for all elements with the exception of O, where each data point covers two bins. \textit{References:} \citealt{Skuladottir15a}~(S); \citealt{North12}~(Mn); \citealt{Skuladottir17}~(Zn).
}
\label{fig:mgplot}
\end{figure*}

\subsection{The relative contributions of massive stars, SN type~Ia, and AGB stars to chemical elements}

For a quantitative comparison with theoretical nucleosynthetic yields, accurate NLTE (and preferentially 3D NLTE) abundances are required, as well as detailed calculations and/or models. However, with the data presented here we are able to make a qualitative evaluation of the relative contribution of different nucleosynthetic sites for each element in Sculptor. For our discussion we make four simplifying assumptions:

\begin{itemize}
\item[1)] SN type Ia and type II (and other core-collapse SN) are the main producers of the $\alpha$- and iron-peak elements.
\item[2)] Mg is predominantly produced by SN type II, and the contribution from SN type Ia is negligible.
\item[3)] For the main stellar population in Sculptor, the contribution of SN type Ia and the $s$-process is negligible at $\text{[Fe/H]}<-2$.
\item [4)]For the elements discussed here, the SN type II yields and 3D NLTE corrections are not strongly metallicity dependent in the range $-2\lesssim\text{[Fe/H]}\lesssim-1$.
\end{itemize}
The first two assumptions are generally accepted, and supported both by theory and observations (e.g. \citealt{Tsujimoto95,Iwamoto99,Kobayashi06,Nomoto13}). Furthermore, the second one is also supported by our own data, as [Mg/Fe] shows the steepest negative slope with [Fe/H] (see Fig.~\ref{fig:alpha}), indicating very little contribution from SN type Ia. The third assumption is safely adopted as there is no evidence of SN type~Ia nor the $s$-process in the measured abundance ratios in Figs. \ref{fig:odd}-\ref{fig:heavy}, at the lowest metallicities. The last assumption is justified in part by theoretical yields, which do not predict a strong metallicity dependence for the elements in question at these metallicities \citep{Kobayashi06}. In addition,  observations in the Milky Way generally show flat trends of [X/Fe] in the range $-2\lesssim\text{[Fe/H]}\lesssim-1$ (see Figs. \ref{fig:alpha} and \ref{fig:iron}), where SNe type II are believed to dominate the metal production, thus making strongly metallicity dependent yields or NLTE-effects unlikely overall. In a few cases, assumption 4) might not hold completely, due to sensitivity to 3D and/or NLTE effects, but for the majority of elements this should be a reasonable approximation for our purposes. 

The abundance measurements shown in Fig. \ref{fig:alpha}-\ref{fig:heavy} and listed in Table~\ref{table:abund} are divided into bins in [Fe/H] and the average of [X/Mg] in each bin are shown in Fig.~\ref{fig:mgplot}. From our assumptions it directly follows that the abundance ratios at the lowest metallicities, $\text{[Fe/H]}<-2$, in Fig.~\ref{fig:mgplot} are direct measurements of the SN type~II contribution to each element in question. Furthermore, if SNe type~II were the only source of metals in Sculptor, all elements would show completely flat trend of [X/Mg] with [Fe/H]. Any increase in [X/Mg] with [Fe/H] therefore indicates a contribution from other sources (which do not affect the Mg abundance). In the case of $\alpha$- and iron-peak elements this is indicative of contributions from SN type~Ia, while for the heavy elements this shows the effects of the $s$- and/or $r$-processes.

\subsubsection{$\alpha$-elements}

In the left panel of Fig.~\ref{fig:mgplot}, the [$\alpha$/Mg] ratios are shown for O, Ca and Ti from this work, as well as NLTE-corrected S abundances from \citet{Skuladottir15a}. In the case of O, stars are binned in two bins instead of four because of lack of data (see Fig.~\ref{fig:alpha}). As already shown in \citet{Skuladottir15a,Skuladottir18}, the $\alpha$-elements in Sculptor do not all have constant ratios with respect to each other as a function of [Fe/H]. The only element with a flat trend is [O/Mg], indicating that the contribution to O by SN type Ia is negligible. For S, Ca, and Ti, on the other hand, the [$\alpha$/Mg] increases with added SN type Ia contribution in Sculptor. This is consistent with theoretical SN type~Ia yields which predict almost no O, but non-negligible yields of the heavier $\alpha$-elements (e.g.\citealt{Tsujimoto95,Iwamoto99}).

\begin{figure}
\centering
\includegraphics[width=\hsize-0.2cm,clip=]{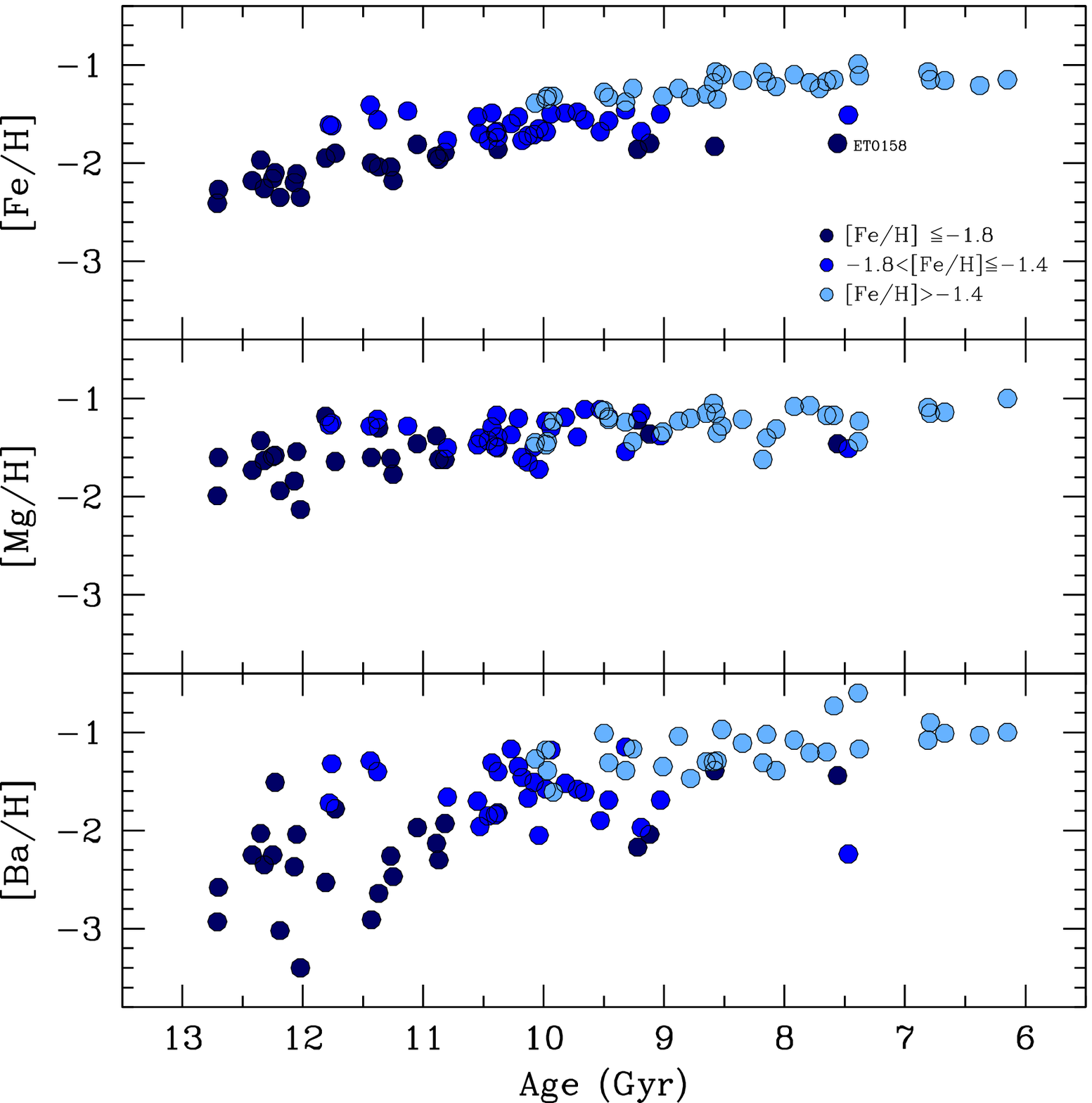}
\caption{Abundances of Sculptor stars as a function of age for [Fe/H], [Mg/H] and [Ba/H]. Colours indicate the metallicity: most Fe-poor stars, $\text{[Fe/H]}\leq-1.8$, are dark blue; blue are $-1.8<\text{[Fe/H]}\leq-1.4$; and most Fe-rich are light blue, $\text{[Fe/H]}>-1.4$.}
\label{fig:ageh}
\end{figure}

\begin{figure}
\centering
\includegraphics[width=\hsize-0.2cm,clip=]{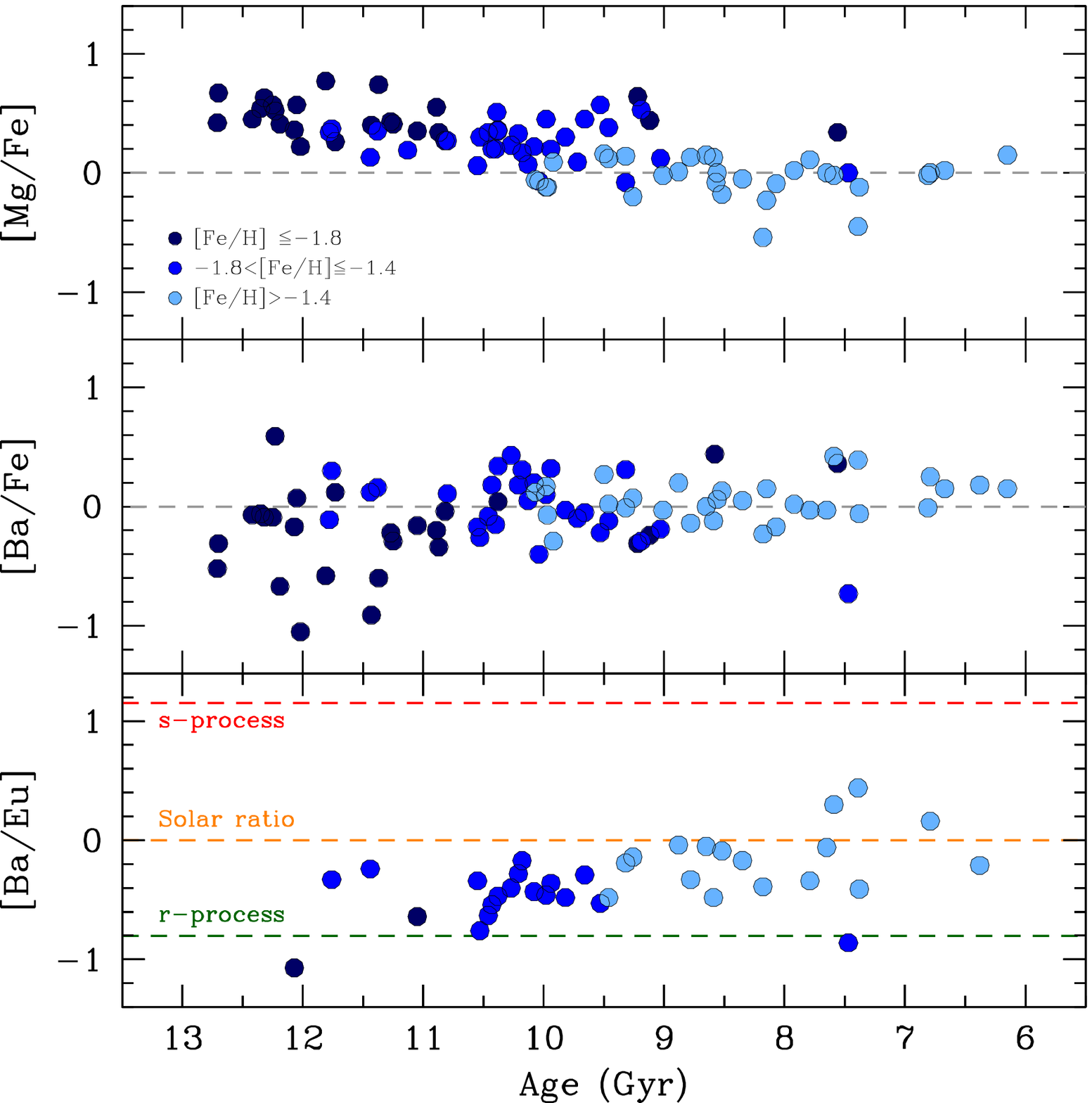}
\caption{Abundance ratios, [Mg/Fe], [Ba/Fe], and [Eu/Ba] in Sculptor stars as a function of age. Red, orange, and green dashed lines in the bottom panel show the value of the $s$-process, the solar ratio, and $r$-process respectively \citep{Bisterzo14}. Blue colours are the same as in Fig.~\ref{fig:ageh}.
}
\label{fig:ageratios}
\end{figure}

\subsubsection{Iron-peak elements}

The iron-peak elements: Sc, Cr, Fe, Co and Ni from this work, and Mn from \citet{North12}, and Zn from our UVES data and \citet{Skuladottir17}; are shown in the middle panel of Fig.~\ref{fig:mgplot}. The elements Fe and Cr show very steeply increasing slopes with [Fe/H], indicating a very efficient production of these elements by SN type Ia. The same is true for Mn, with the possible exception of the most metal-poor data point, which only includes 3 stars, as measuring Mn becomes challenging at these metallicities. However, overall, the slopes of Mn, Fe, and Cr are very similar, and both [Cr/Fe] and [Mn/Fe] show a fairly flat trend with [Fe/H] (see Fig.~\ref{fig:iron} and \citealt{North12}). Thus, [Mn/Fe] and [Cr/Fe], seem to be very similar in the yields of SN type II and type Ia. This is also seen in the Milky Way, where populations which separate into high- and low-$\alpha$ (and thus smaller and larger SN type Ia contributions) at similar metallicities ($-1.6\lesssim\text{[Fe/H]}\lesssim-0.8$) do not show such differences neither in [Cr/Fe] nor [Mn/Fe] (\citealt{NissenSchuster10,NissenSchuster11}). 

The light, odd, iron-peak element Sc also shows a very significant contribution from SN type~Ia. However, there is a clear separation in atomic number, as the elements heavier than iron, Co, Ni, and Zn, all show only moderate or no contribution from SN type Ia. We advise slight caution for interpreting the quantitative trends of Sc and Ni as these elements are only measured from very weak lines, in the regime where distinguishing between upper limits and actual detections becomes challenging. This would result in slightly underestimate the increase of [X/Mg] with [Fe/H], but no drastic changes are expected.

These combined results of the iron-peak elements (in particular the small contribution from Ni) seem to indicate a dominant contribution of low-metallicity SN type Ia of sub-Chandrasekhar-mass explosions of white dwarfs \citep{Sim10,McWilliam18}. In the study of \citet{NissenSchuster10,NissenSchuster11},  a similar relation was found where both Ni and Zn correlated with the high and low $\alpha$-abundances, indicating less contribution from SN type~Ia compared to the lighter iron-peak elements. Unfortunately Co was not included in their study. At higher metallicities in the Milky Way this correlation is less clear in the case of Ni \citep{Mikolaitis17}. But the SN type~II yields of all these elements, Co, Ni and Zn are predicted to be quite metallicity dependent at $\text{[Fe/H]}>-1$ \citep{Kobayashi06}, making any interpretation not very straightforward. The case of Zn is discussed in detail in \citet{Skuladottir17}, but both theory and observations are consistent with Zn not being significantly produced in SN type~Ia.

\subsubsection{Heavy neutron-capture elements}

The heavy elements Ba, La and Eu over Mg are shown in the right panel of Fig.~\ref{fig:mgplot}. The sharp increase in [Ba/Mg] with [Fe/H] results from an added production of Ba by the $s$-process, which happens on similar timescales to SN type~Ia. The element Eu is mainly produced by the $r$-process, however, the lack of data at the lowest metallicities prevents us from drawing strong conclusions regarding Eu at the earliest epochs. At the higher metallicities in Sculptor, $\text{[Fe/H]}> -2$, [Eu/Mg] is constant. This does not necessarily mean that there is no increase in both these elements. Indeed, from metallicities $-2< \text{[Fe/H]} < -1.6$ to $\text{[Fe/H]} > -1.2$ there is an increase of $0.19\pm0.11$~dex in [Eu/H], and a comparable increase in [Mg/H] of $0.24\pm0.08$. The constant value of [Eu/Mg], however, excludes any significant time delay in the production of Eu compared to Mg on the order of $\gtrsim$1 Gyr. 

The heavy element La is predicted to be produced both in the $s$- and $r$-processes \citep[75\% of the solar La is contributed by the $s$-process, according to][]{Bisterzo14}. However, the contribution from the $s$-process is not obvious from Fig.~\ref{fig:mgplot}. The most metal-poor [La/Mg] point only includes 5 stars, and is furthermore likely to be biased towards higher values, as the line is only measurable when the La value is high. But when the three metal-rich bins are considered, there is no clear trend of added contribution from the $s$-process to this element. However, these La abundances come from very weak lines so the mean [La/Mg] could be overestimated at the lowest metallicities. The details of the trends of the heavy neutron capture elements, along with new measurements for this stellar sample will be discussed in more detail in Sk\'{u}lad\'{o}ttir et al. in prep.

\subsection{Evidence for top-light IMF?}

The Sagittarius dSph galaxy has been shown to be more deficient in hydrostatic $\alpha$-elements (O, Mg), compared to the explosive $\alpha$-elements (Si, Ca, Ti) at the highest metallicities, i.e. $\text{[Ca/Mg]}\approx+0.2$ \citep{McWilliam13,Hasselquist17}. In addition, very high abundances of $\text{[Eu/Mg]}_{r}\approx+0.4$ are observed (when Eu has been corrected for contribution from the $s$-process). The authors suggest that these combined results could be explained by a top-light initial mass function (IMF) in Sagittarius, missing the most massive supernovae, whose yields are relatively rich in the hydrostatic $\alpha$-elements.

As is shown in Fig.~\ref{fig:mgplot}, these high values of [Ca/Mg] and [Eu/Mg] are not seen in our Sculptor data, even at the highest metallicities. However, there is a clear increasing trend of [S,Ca,Ti/Mg] with increasing [Fe/H]. This trend is difficult to explain with a top-light IMF as it requires either the IMF to change over time, or the yields of SN type~II to be very metallicity dependent, which is not seen in the Milky Way at the same metallicities (see Fig.~\ref{fig:alpha}). At the lowest metallicities in Sculptor, $\text{[Fe/H]}<-2$ where the contribution from SN type Ia is negligible, all abundance ratios [$\alpha$/Fe] are consistent with what is observed in the Milky Way halo, showing no signs of differences in the IMF. Finally we note that an increase of e.g. [S/O] with [Fe/H] is also seen in the Milky Way disk at $\text{[Fe/H]}>-1$ where the contribution from supernovae type~Ia is significant (see \citealt{Skuladottir18} (Appendix) and references therein), thus making the origin of this trend very clear. We therefore conclude that our data shows no convincing evidence for a top-light IMF in the Sculptor dSph.

Alternatively, instead of different IMF, it is not clear how well the IMF is sampled in a small system with a low star formation rate (e.g. \citealt{Tolstoy03}). However, if Sculptor is suffering from incomplete IMF sampling, the effect is small enough to be hidden within our measurement uncertainties. Smaller systems could a better place to witness the effects of incomplete IMF sampling.

\subsection{Timescales in Sculptor}

A detailed star formation history for Sculptor was derived from deep CMD covering the whole spatial extent of the galaxy \citep{deBoer12}, while using the observed metallicity distribution of Sculptor as a constraint \citep[][]{Battaglia08b,Kirby11,Starkenburg10}. In this work, \citet{deBoer12} also selected isochrones to derive ages of individual RGB stars, using their magnitude, colour, [Fe/H] and [$\alpha$/Fe]. This technique results in relatively low uncertainties on the ages,  $\Delta\text{age}\approx\pm1.8$~Gyr on average for our sample.

The elemental abundances of Fe, Mg, and Ba in Sculptor all have differences in their behaviour with age, as shown in Fig.~\ref{fig:ageh}.\footnote{The star ET0158 is Li-enhanced compared to the rest of the sample, and is possibly an AGB star (see Section~5.1). This star has a rather blue colour (see Fig.~\ref{fig:cmd}) resulting in an unusually young age for its metallicity (see Fig.~\ref{fig:ageh}).} The slope of [Fe/H] with age is steeper compared to that of [Mg/H], as it traces both the contribution of type~II and type~Ia SNe, while the increase in Mg with time is only contributed by SN type~II (e.g. \citealt{Iwamoto99,Kobayashi06,Nomoto13}). 
The neutron-capture element Ba shows a very large scatter at the earliest times, $>$11\,Gyr ago, exceeding both measurement uncertainties, and that of Fe and Mg. But around $\sim$11\,Gyr ago, the main $s$-process became dominant and the scatter decreased. The large scatter at the earliest epoch is associated to the $r$-process, which does not trace normal SN type II production of $\alpha$-elements or iron at these times/metallicites (see Sec.~\ref{sec:rs_trends}).

The abundance ratios [Mg/Fe], [Ba/Fe] and [Ba/Eu] with age are depicted in Fig.~\ref{fig:ageratios}. As shown in \citet{deBoer12}, [Mg/Fe] has a well defined decreasing trend with age, consistent with the contribution of SN type Ia becoming more important with time. Similar to [Ba/H], [Ba/Fe] has a large scatter at the earliest times. The bottom panel of Fig.~\ref{fig:ageratios} shows the relative ratios of the $s$- and $r$-processes with the [Ba/Eu] ratio. Similar to what is shown in Fig.~\ref{fig:baeu}, [Ba/Eu] at the earliest times is consistent with the $r$-process dominating the production of the heavy elements in Sculptor, but as time passes, the $s$-process becomes more significant. 

Comparing all three panels, as well as Fig.~\ref{fig:alpha}, \ref{fig:heavy} and \ref{fig:baeu}, seems to indicate that the timescales of the $s$-process and SN type~Ia in Sculptor are comparable. The current dearth of Eu measurements at the oldest ages however prevent us from dating precisely the onset of the $s$-process. The fact that the high [Ba/Fe] dispersion diminishes at the same time as the [Ba/Fe] reaches the solar value, $\sim$11\,Gyr ago is probably the best trace of the $s-$process onset.

\section{Conclusions}

We have analysed high-resolution VLT/GIRAFFE and VLT/UVES spectra of 99 red giant branch stars in the Sculptor dwarf spheroidal galaxy, to measure the abundances of $17$ chemical elements made up by different nucleosynthetic channels: Li, Na, 
$\alpha$-elements (O, Mg, Si, Ca Ti), iron-peak elements (Sc, Cr, Fe, Co, Ni, Zn), and $r$- and $s$-process elements (Ba, La, Nd, Eu). The sample stars have a wide range in metallicity, $-2.3 <$ [Fe/H] $< -0.9$, populating the whole metallicity distribution of the galaxy with the exception of the very low-metallicity tail, which has been studied elsewhere \citep{Tafelmeyer10,Frebel10,Starkenburg13,Jablonka15,Simon15}. 
Armed with these high-precision elemental abundances we have examined the details of how metal-enrichment proceeds in a small galaxy with a single-peaked star formation which lasted several Gyr before fading away. In many ways, the abundance ratios evolve with metallicity and with time according to expectations from basic nucleosynthetic prescriptions, making Sculptor a textbook galaxy to study chemical evolution.

Our dataset establishes with reasonably good precision, and on statistical grounds, a number of chemical properties of Sculptor that stem naturally from the star formation history of this system:
\begin{itemize}
\item There is a marked decrease in [$\alpha$/Fe], which starts at the Galactic halo plateau value for low [Fe/H] and decreases steadily after a knee, to sub-solar \afe for high [Fe/H], in agreement with expectations given the star formation history of Sculptor, with a dominance of the products of massive stars dying as core-collapse supernovae at early times, and an onset of SN~Ia as early as $\sim$12\,Gyr ago.
\item The position of the knee, around $\text{[Fe/H]}\approx-1.8$, occurs at much lower metallicity than in the Milky Way disks \citep[e.g.][]{Bensby14} or bulge \citep[e.g.][]{Hill11,Gonzalez11}, in agreement with the lower star formation efficiency in Sculptor. 
The position of the knee can also be affected by the ability of the galaxy to retain freshly formed metals \citep[e.g.][]{Lanfranchi07,Vinc16,Cote17}, and more generally the origin, state and fate of the gas in these galaxies, where feedback plays a role not only in regulating star formation but also gas enrichment \citep[e.g.][]{RevazJablonka12}.
\item The products from low-mass AGB stars, as traced by the $s$-process, are also incorporated in the chemical evolution of Sculptor on a longer timescale than massive stars, more similar to that of SN type~Ia.
\item  Except for neutron-capture elements in the early phases of the evolution, the scatter around mean trends in Sculptor for [Fe/H]$>-2.3$ is extremely low, compatible with observational errors. In addition, there is little evidence for scatter in the age-metallicity relations. This calls for an efficient mixing of metals in the gas at all times, at least in the last $\sim$12\,Gyr. 
This has inspired modes that include the mixing \citep{RevazJablonka18} or diffusion \citep[e.g.][]{Escala18} of metals in the galaxy.
\end{itemize}

As the origin of Sculptor chemical enrichment is quite straightforward, we can also refine the empirical constraints on nucleosynthesis processes:
\begin{itemize}
\item We estimated the relative importance of SN~Ia contributions to individual iron-peak and $\alpha$ elements. The most important contribution of SN type~Ia is to the iron-peak elements: Fe, Cr and Mn; however, there is also a modest but non-negligible contribution to both the heavier $\alpha$-elements: S, Ca and Ti, and some of the iron-peak elements: Sc and Co. In Sculptor the lightest $\alpha$-elements (O and Mg) and the heaviest iron-peak elements (Ni and Zn) are consistent with having little or no contribution from SN type~Ia.
\item Sculptor also sheds light on the production of neutron-capture elements through the $r$-process channel, showing a gradual and regular enrichment by the main $r$-process all along the metallicity range, at odds with the idea that the $r$-process in dwarf galaxies would come from very rare events polluting gas to very high levels. This has been seen widely in classical dwarf spheroidal galaxies (e.g. \citealt{Letarte10,Lemasle12,Lemasle14}), whereas smaller systems such as the ultra-faint dwarf galaxies seem to populate more extreme $r$- enhancements (e.g. Ret\,II, \citealt{Ji16Nat}) or very low levels of $r$-products \citep[e.g.][]{Frebel14,Ji16boo}.
\item The chemical evolution in Sculptor does not show signs of having significantly different initial mass function compared to the Milky Way.
\end{itemize}

The very earliest days of the Sculptor galaxy, however, are still poorly represented in our sample, in part because this sample was drawn from the inner 25' radius of the system whose tidal radius is approximately 77' \citep{Mateo:1998}, which badly samples the oldest and most metal-poor population of Sculptor \citep{Tolstoy04,Colemanetal:2005,Clementini05, Battaglia08b,deBoer12}. Obtaining the detailed abundances of sizeable samples of red giants in Sculptor with $\text{[Fe/H]}<-2$, in particular in the range $-3<\text{[Fe/H]}<-2$, has the power to shed light on a variety of open questions on the earliest epochs of star formation, when [Fe/H] was possibly not yet a good proxy for time. This will be key to answering many open questions: the $r$-process production site and dispersion mechanism, the first traces of the $s$-process, nucleosynthesis -or the absence thereof- of massive stars in dwarf galaxies, the first star formation in a small dark matter halo and its relation to C-enrichment, among others.

\begin{acknowledgements}
\'{A}.S. acknowledges funds from the Alexander von Humboldt Foundation in the framework of the Sofja Kovalevskaja Award endowed by the Federal Ministry of Education and Research.
G.B. gratefully acknowledges financial support by the Spanish Ministry of Economy and Competitiveness (MINECO) under the Ramon y Cajal Programme (RYC-2012-11537) and the grant AYA2014-56795-P. 
E.S. gratefully acknowledges funding by the Emmy Noether programme from the Deutsche Forschungsgemeinschaft (DFG).
T.d.B. acknowledges support from the European Research Council (ERC StG-335936).
\end{acknowledgements}
graphystyle{aa}
\bibliography{dsph,mphs,mw,general,lithium,et,vanessa}
\clearpage
\pagebreak


\begin{appendix}

\section{Verification of the abundance analyses}

To ensure the accuracy of our abundance results from FLAMES/GIRAFFE spectra, six stars with detailed abundance determinations from UVES slit spectroscopy were reobserved: Scl-459, Scl-461, and Scl-482 from \citet{Shetrone03}, hereafter S+03;
and Scl-195, Scl-770, and Scl-1446 from
\citet{Geisler05}, hereafter G+05. In addition, the star ET0097 was discovered to be a carbon-enhanced metal-poor (CEMP-no) star, and thus reobserved by \citet{Skuladottir15b}, hereafter Sk+15b, with UVES slit spectroscopy. The nature of the comparison with ET0097 is slightly different, as our GIRAFFE analysis does not account for CN molecular lines, which are present in the HR UVES spectra. The results for these stars, as well as their names for cross-identification, are listed in Table~\ref{table:comp}.

\subsection{Equivalent width (EW) measurements}

The FLAMES/GIRAFFE line strengths were determined from DAOspec, using the line list given in Table~\ref{table:linelist}. The FWHM given by DAOspec was consistent over all HR settings, see Table~\ref{table:fwhm}, as is expected for an RGB stellar sample at this resolution. The higher $\sigma$ in the HR15 setting results from lower number of available lines in this region. While DAOspec fits only Gaussian profiles to the spectral lines, the lower resolution of the GIRAFFE spectra results in line profiles that have an instrumental broadening that is well matched to a Gaussian over our EW measurement range (EW $\le$ 300 m\AA).

The method used here is different from S+03 and G+05, where lines were measured individually using \textsl{splot} in IRAF, but the agreement in EW measurements is overall good. There were slight systematic offsets between this work and S+03, which can be traced directly to the S/N ratio of the spectra, and thus the continuum placement. For high S/N ratio spectra, the continuum level is clear; but as the S/N ratio lowers, DAOspec favours the centre of the noise, whereas the method adopted by S+03 sets the continuum slightly higher, at 2/3 the noise. Thus, the DAOspec EWs might be slightly and systematically smaller in low S/N ratio spectra. Our comparisons imply, however, that these offsets are within our EW measurement error estimates (worst case offset is $\leq4$~m\AA\ while the minimum adopted EW error is 6~m\AA). Detailed comparisons with the G+05 data shows some EWs which are significantly stronger than our FLAMES/GIRAFFE  measurements, particularly for ET0051/Scl-1446.   Each of these lines has been examined in our GIRAFFE spectra for unrecognised blends, but none have been found. Overall, however, the agreement with S+03 and G+05 is within the adopted measurement errors.

\begin{table*}

\centering
\caption{Comparisons of the derived stellar parameters and [Fe/H] with previously published UVES results.\label{table:comp}}
\vspace{0.3cm}
\footnotesize
\tabcolsep=0.105cm
\begin{tabular}{l|ccccc|cccccl}
\hline\hline
Names  & \teff  &  \logg & \vt & [Fe\,I/H] & [Fe\,II/H] & \teff  &  \logg & \vt &  [Fe\,I/H] & [Fe\,II/H] & Ref. \\
 & K  &  & km s$^{-1}$ &   &  & K  &   & km s$^{-1}$ &  &  & \\
\hline		
ET0071/Scl-482  & 4243 & 0.5 & 1.7 & $-1.35\pm0.14$ & $-1.27\pm0.16$ & 4400 & 1.10 & 1.70 & $-1.24\pm0.07$  & $-1.26\pm0.12$ & \citealt{Shetrone03} \\
ET0151/Scl-461  & 4281 & 0.6 & 1.7 & $-1.77\pm0.16$ & $-1.70\pm0.14$ & 4500 & 1.20 & 1.70 & $-1.56\pm0.07$  & $-1.58\pm0.12$ & \citealt{Shetrone03} \\
ET0389/Scl-459  & 4394 & 0.8 & 1.5 & $-1.60\pm0.22$ & $-1.47\pm0.20$ & 4500 & 1.00 & 1.65 & $-1.66\pm0.07$  & $-1.65\pm0.12$ & \citealt{Shetrone03} \\
ET0051/Scl-1446 & 3971 & 0.5 & 1.7 & $-0.92\pm0.12$ & $-0.80\pm0.33$ & 3900 & 0.00 & 2.30 & $-1.20\pm0.13$  & & \citealt{Geisler05}\\ 
ET0113/Scl-195  & 4285 & 0.2 & 1.8 & $-2.18\pm0.19$ & $-2.09\pm0.14$ & 4250 & 0.20 & 1.80 & $-2.10\pm0.15$  & & \citealt{Geisler05}\\
ET0141/Scl-770  & 4188 & 0.3 & 1.9 & $-1.68\pm0.15$ & $-1.62\pm0.13$ & 4075 & 0.00 & 1.90 & $-1.72\pm0.13$  & & \citealt{Geisler05}\\
ET0097  & 4300 & 0.5 & 2.0 & $-1.91\pm0.16$ & $-1.84\pm0.26$ & 4383 & 0.75 & 2.25 & $-2.03\pm0.10$  & $-1.86\pm0.14$ & \citealt{Skuladottir15b}\\
\hline
\end{tabular}
\end{table*}

\subsection{Stellar parameters}

\begin{table}

\centering
\caption{The FWHM (mean and $\sigma$, per pixel) for the FLAMES/GIRAFFE settings, as determined by DAOspec ($1~\text{pixel}=0.05$~\AA).\label{table:fwhm}}
\vspace{0.2cm}
\tabcolsep=0.11cm
\begin{tabular}{lcc}
\hline\hline
Setting & FWHM & $\sigma$ \\
\hline		
HR10 & 6.10 & 0.27\\
HR13 & 6.35 & 0.27\\
HR14 & 5.69 & 0.45\\
HR15 & 6.70 & 1.16\\
\hline
\end{tabular}
\end{table}

The effective temperatures, \teff, were determined differently in S+03, G+05, and Sk+15b, and each differ from this analysis as well. S+03 used \ion{Fe}{i} as a function of excitation potential, $\chi$, with a starting point based on (B-V) dereddened colour and assuming a metallicity from the star's CMD location. G+05 adopted temperatures from (V-K) and (J-K) colours based on calibrations from \citet{Bessell98}. Sk+15b adopted photometric temperatures based on \citet{RamirezMelendez05}. S+03 and G+05 determined gravity, \logg, using the same method as here, but ionisation equilibrium delivers gravities that are dependent on the adopted \teff. In Sk+15b photometric gravities were used. Microturbulence velocities, \vt, were determined in all three papers by minimising the \ion{Fe}{I} abundance dependency on observed equivalent widths, whereas here the expected line strength was used.


\begin{figure}

\centering
\includegraphics[width=\hsize,clip=]{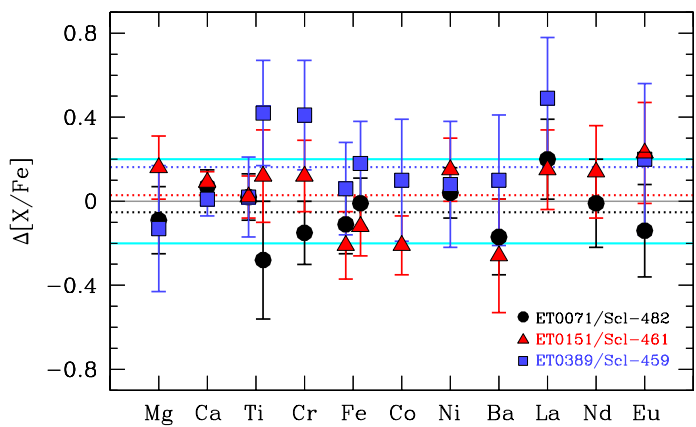}
\caption{
Abundance differences between our GIRAFFE analysis and S+03, $\Delta\text{[X/Fe]}=\text{[X/Fe]}_\text{GIR} - \text{[X/Fe]} _\text{S+03}$. In the case of Fe, $\Delta\text{[Fe/H]}$ is plotted instead. For Ti and Fe, the ionised species are plotted to the right. Dotted lines are the average abundance offsets for each star, while solid cyan lines show the expected GIRAFFE precision.\label{fig:compS03}
}
\end{figure}

\begin{figure}

\centering
\includegraphics[width=\hsize]{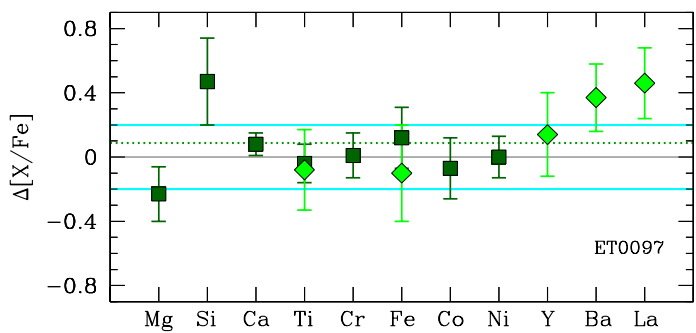}
\caption{Same comparison as in Fig.~\ref{fig:compS03} for ET0097 in Sk+15b. Neutral species are plotted with dark green squares, ionised with green diamonds. Lines are the same as in Fig~\ref{fig:compS03}.\label{fig:comp97}
}
\end{figure}

A comparison of the atmospheric parameters for the seven stars, which overlap with S+03, G+05 and Sk+15b, is shown in Table~\ref{table:comp}. Temperatures in this work are slightly cooler than those determined by S+03 ($\Delta \text{\teff}=-161 \pm42$~K), and warmer for the G+05 stars ($\Delta \text{\teff}=+73 \pm32$~K). Accordingly, our gravities are lower than in S+03 ($\Delta \text{\logg}=-0.47 \pm0.19$) and higher than in G+05  ($\Delta \text{\logg}=+0.27 \pm0.20$). Microturbulence velocities agree within error bars, except for the coolest star of the G+05 sample where we find a microturbulence 0.5~km.s$^{-1}$ lower. In fact, for this cool star, the GIRAFFE spectra did not allow us to constrain \vt\ satisfactorily, which is reflected in a larger metallicity error. The only significant differences are therefore the slightly hotter S+03 temperature scale (and its correspondingly higher gravities). However, the resulting metallicities are compatible with those of S+03 and G+05 within the errors. For ET0097, \teff, \logg, \vt\, and metallicity all agree within errorbars between our analysis and Sk+15b, indicating that the CN molecular lines were not a serious issue for this star at the resolution and wavelength range observed in this work.

Finally, DAOspec was used to measure all the spectral lines in the UVES spectra of S+03 to directly compare the different analyses. The agreement is very good over most of the EW range; however, the strongest lines are not well fit by Gaussian profiles at the resolution of UVES, and thus DAOspec underestimates the EWs by not fitting the damping wings.  This is not a significant problem for the FLAMES/GIRAFFE spectra since the resolution is lower, and therefore the profile is dominated by the instrumental (gaussian) profile all the way to at least $\sim$200m\AA. More problematic is that some stars show a trend of increasingly disparate measurements with increasing line strength. This is more concerning because of the spectroscopic methods used to determine the stellar atmospheric parameters.  
For example, a trend of decreasing EW and thus decreasing resulting abundance in \ion{Fe}{i} lines will mimic a higher microturbulence value, resulting in incorrect stellar parameters. Therefore, we conclude that at the higher resolution of the UVES spectra, DAOspec should not be used for spectral line measurements (at least above EW$\sim$120m\AA), however, it seems to be very well matched to the resolution of the FLAMES/GIRAFFE spectra.

\subsection{Abundance trend}

There are some differences in the stellar parameters adopted here compared to those of S+03, G+05, and Sk+15b (see Table~\ref{table:comp}), and the same is true for the model atmospheres, since all previous analyses used MARCS models \citep{Gustafsson75}. 
The linelist and atomic data used here (see Table~\ref{table:linelist}) also differs slightly from that of S+03 and G+05, and significantly from Sk+15b. Finally, S+03 and G+05 used the LTE spectrum synthesis code MOOG \citep{Sneden73} for abundance determinations, and in Sk+15b, the LTE code Turbospectrum \citep{Plez12} was used, while here we use a different code, originally developed by \citet{Spite67}. 

To test how these slight differences affect the results, our FLAMES/GIRAFFE abundance measurements are compared to S+03 in Fig.~\ref{fig:compS03}. The agreement is excellent for most elements in all three stars. The elements with more discrepancy include \ion{Ti}{II} and \ion{La}{II}, which are both detected with very few (1-3) and very weak lines in the GIRAFFE spectra. Despite an offset in the Fe abundances in ET0151/Scl-461 (due to a 200~K difference in \teff\ and 0.6 dex in \logg), most relative element abundances are still in good agreement with S+03. The star ET0389/Scl-459 has the lowest S/N of the three (and among the lowest in the GIRAFFE sample) but still shows a rather good agreement for most species (10 of 13 elements) in common with S+03. Those in poor agreement are elements measured from weak lines, and typically have higher abundances from our analysis. This is the only star for which the overall abundance pattern is displaced, $<\Delta\text{[X/Fe]}>=+0.14$ over 13 elements.

The comparison of the CEMP-no star, ET0097, between this analysis and Sk+15b is shown in Fig.~\ref{fig:comp97}. The linelist used in Sk+15b is significantly different from the one used here, in particular as the UVES spectrum is missing the wavelength coverage offered by HR10, but includes both bluer and redder regions than available in our GIRAFFE spectra. Overall the agreement is excellent, with few notable exceptions. Like discussed in Section~4.3.2, the \ion{Si}{I} line at 6244.5~\AA, which is used for ET0097 in the GIRAFFE sample, gives systematically high Si abundances, with an offset of $\sim+0.30$\,dex, explaining the discrepancy seen in Fig.~\ref{fig:comp97}. Out of the two \ion{Ba}{II} lines in common with both analyses, 6141.7 and 6497.0~\AA, the redder line is in agreement, while a significantly lower abundance is derived by Sk+15b for the bluer line. For Mg and La there are no overlapping lines used in these analyses, as stronger \ion{La}{II} lines in the blue of the UVES spectrum are favoured. In the case of the GIRAFFE spectrum, La is determined from 2 very weak lines ($<30$~ m\AA), close to our detection limit.

\section{Comparison with intermediate resolution spectroscopy}

\begin{figure}
\centering
\includegraphics[width=\hsize-0cm,clip=]{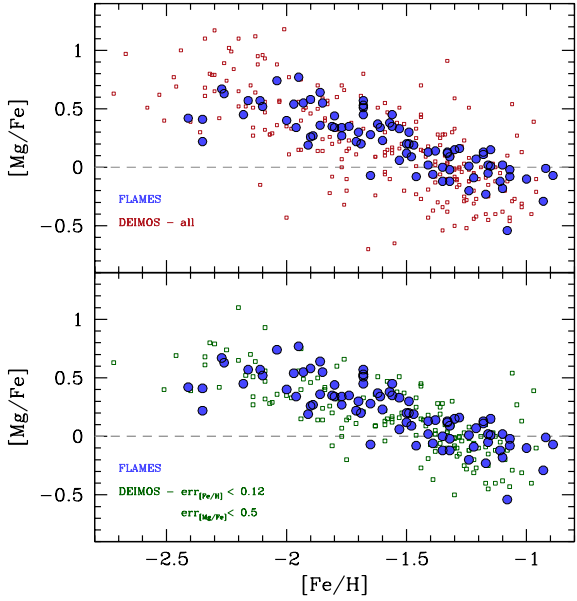}
\caption{
Comparison of [Mg/Fe] between our HR FLAMES results (blue cirles) and those of Keck DEIMOS (squares). Top panel shows all published measurements by \citet{Kirby09,Kirby11} in red, while the bottom panel shows only the most reliable measurements, ($\text{err}_\text{[Fe/H]}<0.12$, and $\text{err}_\text{[Mg/Fe]}<0.5$) in green.
}\label{kirbycomp}
\end{figure}


At present there is a large number of intermediate resolution, R~$\sim~5000-10000$, multi-object spectrographs that allow us to
obtain detailed spectra of individual stars in nearby galaxies. One particular approach has made use of the entire extensive forest of small lines over a large wavelength range without significant individual detections, which are then compared to synthetic spectra rather than being measured individually (e.g \citealt{Kirby08}). 

The abundances of a range of elements (Fe, Mg, Ca, Si and Ti) were determined in this way for a large sample of 376 RGB stars in the centre of the Sculptor dSph using Keck DEIMOS spectra \citep{Kirby09, Kirby11}. These lower resolution abundance measurements are compared to our HR FLAMES analysis in Fig.~\ref{kirbycomp}. As shown, the method applied by \citet{Kirby09, Kirby11} can efficiently obtain a large sample of abundances, and shows the general trends in $\alpha$-element abundance ratios. Predictably, however, the precision is less than if individual lines can be accurately measured with HR and high S/N spectra. This is therefore not the best method to understand the scatter in abundance ratios, and, as we show here, it also may suffer from inaccuracies which increase with decreasing metallicity as many of the weak lines disappear.

From Fig.~\ref{kirbycomp} it is evident that the overall picture changes if only the most reliable points from \citet{Kirby09, Kirby11} are plotted. If all measurements are included, there is no change in the slope, so-called knee, in the [Mg/Fe] vs [Fe/H] in Sculptor, suggesting that Sculptor has had very early enrichment by SN Ia \citep{Kirby11}. However, if only measurements with $\text{err}_\text{[Fe/H]}<0.12$ and $\text{err}_\text{[Mg/Fe]}<0.5$ are shown, the results become more consistent with present work, i.e. showing a knee in the distribution of measurements at $\rm[Fe/H]< -1.8$, which we interpret as the time when SN~Ia start to contribute to the chemical enrichment, $\sim$1-2~Gyr after the onset of star formation \citep{deBoer12}.

\section{Online tables}

\clearpage
\begin{deluxetable}{lrrrrrrrcrrrl}
\center
\tabletypesize{\tiny}
\scriptsize
\tabcolsep=0.03cm
\tablecaption{\label{table:PhotVr} GIRAFFE and UVES Target Stars in Sculptor. Stars considered non-members (NM) are noted, as well as stars with a different membership definition by Gaia.}
\tablewidth{0pt}
\tablehead{
\colhead{Target}  & \colhead{RA}  &  \colhead{DEC}  
  & \colhead{V}  &  \colhead{I}  &  \colhead{J}& \colhead{K}  
  & \colhead{v$_r$}  &   \colhead{S/N} &\colhead{Gaia DR2 ID} &\colhead{RA PM$\pm$err} &\colhead{Dec PM$\pm$err} &\colhead{Comment}\\
\colhead{}  & \colhead{J2000}  &  \colhead{J2000}  
  & \colhead{mag}  &  \colhead{mag}  &  \colhead{mag}& \colhead{mag}  
  & \colhead{km/s}  &   \colhead{} &   \colhead{} &   \colhead{mas/yr} &   \colhead{mas/yr} & \colhead{}
}
\startdata

ET0009  &   1 00 54.18 & -33 40 14.60 & 17.149 & 15.835  & 14.743 & 13.882 &   95.77 $\pm$0.27 &   33  &    5027216583830257536  &    0.074 $\pm$0.100 &    -0.166 $\pm$0.090         &\\                  
ET0013  &   1 00 50.24 & -33 36 38.20 & 17.335 & 16.121  &    ... &    ... &  105.07 $\pm$0.95 &   22  &    5027217820780833536  &    0.188 $\pm$0.099 &     0.070 $\pm$0.099         &\\                  
ET0024  &   1 00 34.04 & -33 39 04.61 & 17.200 & 15.654  &    ... &    ... &  112.88 $\pm$0.37 &   40  &    5027220187306194048  &    0.217 $\pm$0.090 &    -0.095 $\pm$0.081         & Gaia-NM\\                  
ET0026  &   1 00 12.76 & -33 41 16.00 & 17.280 & 16.025  & 14.966 & 14.151 &   98.36 $\pm$0.30 &   51  &    5027218714133971712  &    0.022 $\pm$0.103 &    -0.178 $\pm$0.088         &\\                                
ET0027  &   1 00 15.37 & -33 39 06.20 & 17.287 & 15.988  & 14.891 & 14.065 &  112.34 $\pm$0.18 &   51  &    5027219676206641152  &    0.201 $\pm$0.096 &    -0.070 $\pm$0.093         &\\                  
ET0028  &   1 00 17.77 & -33 35 59.71 & 17.345 & 16.002  &    ... &    ... &  120.24 $\pm$0.22 &   43  &    5027221359833813504  &    0.076 $\pm$0.094 &     0.004 $\pm$0.090         &\\                  
ET0031  &   1 00 07.57 & -33 37 03.89 & 17.457 & 16.262  & 15.239 & 14.508 &  115.05 $\pm$0.27 &   38  &    5027220019804016768  &    0.065 $\pm$0.112 &    -0.119 $\pm$0.120         &\\                  
ET0033  &   1 00 20.29 & -33 35 34.51 & 17.495 & 16.295  &    ... &    ... &  108.68 $\pm$0.22 &   38  &    5027221359833858944  &    0.066 $\pm$0.113 &    -0.061 $\pm$0.102         &\\
\enddata
\end{deluxetable}

\clearpage
\begin{deluxetable}{lccrc|lccrc}
\small
\tablecaption{Linelist\label{table:linelist}}
\tablewidth{0pt}
\tablehead{
\colhead{Ion} &\colhead{Wavelength}  & \colhead{$\chi$}  &  \colhead{log gf}  
  & \colhead{Ref.} & \colhead{Ion}&\colhead{Wavelength}  & \colhead{$\chi$}  &  
\colhead{log gf}  & \colhead{Ref.} \\[.2ex]
\colhead{} & \colhead{\AA}    & \colhead{eV}   &\colhead{dex}    &\colhead{~~}  & \colhead{~~} &\colhead{\AA}    & \colhead{eV}   &\colhead{dex}    &\colhead{~~}
}
\startdata
Li I	&	6707.80	&	0.00	&$	0.093	$&	   S03  	&	Mn I	&	5516.77	&	2.18	&$	-1.847	$&	   S03   \\
O I	&	6300.31	&	0.00	&$	-9.750	$&	   S03 	&	Mn I	&	6013.51	&	3.07	&$	-0.252	$&	   S03 	\\
O I	&	6363.79	&	0.02	&$	-10.250$&	   S03 	&	Mn I	&	6021.82	&	3.08	&$	0.035	$&	   S03 	\\
Na I	&	5682.65	&	2.10	&$	-0.700	$&	   S03 	&	Fe I	&	4966.10	&	3.33	&$	-0.890	$&	   S03 	\\
Na I	&	5688.21	&	2.10	&$	-0.370	$&	   S03 	&	Fe I	&	5006.12	&	2.83	&$	-0.628	$&	   S03 	\\
Na I	&	5889.97	&	0.00	&$	0.122	$&	   S03 	&	Fe I	&	5079.75	&	0.99	&$	-3.240	$&	   S03 	\\
Na I	&	6154.23	&	2.10	&$	-1.560	$&	   S03 	&	Fe I	&	5083.35	&	0.96	&$	-2.862	$&	   S03 	\\
Na I	&	6160.75	&	2.10	&$	-1.260	$&	   S03 	&	Fe I	&	5150.85	&	0.99	&$	-3.030	$&	   S03 	\\
\enddata
\tablerefs{S03=Shetrone et al. (2003); P08=Pompeia et al. (2008); L10=Letarte et al. (2010)}
\end{deluxetable}

\clearpage

\begin{deluxetable}{lcccccccccccrrr}
\small
\tabcolsep=0.12cm
\tablecaption{\label{table:Stellpar} Atmospheric Parameters. Effective temperatures, T$_\text{eff}$, are listed for three colours as well as the adopted value. Gravities, $\log$ g, are listed as determined from photometry (phot.) and spectroscopy (spec.), and the latter is adopted.
}
\tablewidth{0pt}
\tablehead{
\colhead{Star}&\colhead{T$_\text{eff}$}    &\colhead{T$_\text{eff}$}    &\colhead{T$_\text{eff}$}  & \colhead{BC$_V$}&\colhead{$M_\textsl{bol}$}& \colhead{$\log$ g}&\colhead{T$_\text{eff}$}    &\colhead{$\log$ g}  &\colhead{$v_t$}        &\colhead{[FeI/H]}&\colhead{[FeII/H]} &\colhead{Age} &\colhead{Age} &\colhead{$\delta$Age}\\[.2ex]
\colhead{}    &\colhead{(V-I)}&\colhead{(V-K)}&\colhead{(V-J)}& \colhead{}      &\colhead{}    & \colhead{phot.}&\colhead{adop.}&\colhead{spec.} & \colhead{spec.}      &\colhead{spec.}       &\colhead{spec.} &\colhead{min}  &\colhead{mean}  &\colhead{error}   \\[.2ex]
\colhead{}    & \colhead{K}   &\colhead{K}    &\colhead{K}  & \colhead{mag}   &\colhead{mag} & \colhead{dex} & \colhead{K}   &\colhead{dex  } &\colhead{km\,s$^{-1}$}&\colhead{dex}   &\colhead{dex}  &\colhead{Gyr} &\colhead{Gyr} &\colhead{Gyr}}  
\startdata
ET0009 	&	4286	&	4105	&	4123	&$	-0.71	$&$	-3.17	$&	0.61	&	4171	&	0.2	&	2.2	&$	-1.68	$&$	-1.62	$&	9.25	&	9.53	&	2.30	\\
ET0013 	&	4416	&	 ... 	&	 ... 	&$	-0.63	$&$	-2.89	$&	0.76	&	4286	&	0.2	&	1.4	&$	-1.68	$&$	-1.55	$&	7.75	&	9.19	&	1.87	\\
ET0024 	&	4027	&	 ... 	&	 ... 	&$	-1.04	$&$	-3.44	$&	0.38	&	3897	&	0.0	&	2.2	&$	-1.24	$&$	-1.07	$&	5.50	&	7.71	&	1.51	\\
ET0026 	&	4361	&	4183	&	4192	&$	-0.65	$&$	-2.98	$&	0.71	&	4245	&	0.5	&	1.7	&$	-1.80	$&$	-1.73	$&	5.50	&	9.12	&	2.01	\\
ET0027 	&	4290	&	4127	&	4116	&$	-0.71	$&$	-3.03	$&	0.66	&	4178	&	0.3	&	2.2	&$	-1.50	$&$	-1.44	$&	7.50	&	9.03	&	1.66	\\
ET0028 	&	4215	&	 ... 	&	 ... 	&$	-0.80	$&$	-3.06	$&	0.61	&	4085	&	0.3	&	2.0	&$	-1.22	$&$	-1.16	$&	6.50	&	8.07	&	1.51	\\
ET0031 	&	4435	&	4291	&	4259	&$	-0.60	$&$	-2.74	$&	0.84	&	4329	&	0.5	&	2.1	&$	-1.68	$&$	-1.61	$&	5.75	&	10.39	&	2.19	\\
ET0033 	&	4432	&	 ... 	&	 ... 	&$	-0.61	$&$	-2.72	$&	0.84	&	4302	&	0.6	&	1.7	&$	-1.77	$&$	-1.70	$&	6.25	&	10.80	&	1.92	\\
\enddata
\tablerefs{Ages come from \citet{deBoer12}}
\end{deluxetable}

\clearpage

\begin{deluxetable}{clcccc}
\small
\tablecaption{\label{table:ew} Equivalent width (EW) measurements. Lines which are not used for the abundance determination are flagged (*). The full table is available online.}
\tablewidth{0pt}
\tablehead{
\colhead{Star}&\colhead{Ion} & \colhead{Wavelength} & \colhead{EW} & \colhead{$\delta$ EW} & \colhead{Flag}  \\[.2ex]
\colhead{}&\colhead{} & \colhead{\AA} & \colhead{m\AA} & \colhead{m\AA} & \colhead{} 
}
\startdata
ET0009	&	O I	&	6300.31	&	...	&	...	&		\\
ET0009	&	Na I	&	6154.23	&	...	&	...	&		\\
ET0009	&	Na I	&	6160.75	&	...	&	...	&		\\
ET0009	&	Mg I	&	5528.41	&	194	&	3.6	&		\\
ET0009	&	Si I	&	6243.82	&	...	&	...	&		\\
ET0009	&	Si I	&	6244.48	&	...	&	...	&		\\
ET0009	&	Ca I	&	6122.23	&	194	&	4.7	&		\\
ET0009	&	Ca I	&	6161.30	&	50	&	5.7	&		\\
\enddata
\end{deluxetable}
\begin{sidewaystable*}
\caption{Chemical abundances.\label{table:abund}}
\vspace{0.2cm}
\tabcolsep=0.04cm
\scriptsize
\begin{tabular}{lrcrcrcrcrcrcrcrcrcrcrcrcrcrcrcrcrcrc}
\hline\hline

Star	&	[Fe/H]	&	$\delta$	&	[O/Fe]	&	$\delta$	&	[Na/Fe]	&	$\delta$	&	[Mg/Fe]	&	$\delta$	&	[Si/Fe]	&	$\delta$	&	[Ca/Fe]	&	$\delta$	&	[Sc/Fe]	&	$\delta$	&	[TiI/Fe]	&	$\delta$	&	[TiII/Fe]	&	$\delta$	&	[Cr/Fe]	&	$\delta$	&	[FeII/Fe]	&	$\delta$	&	[Co/Fe]	&	$\delta$	&	[Ni/Fe]	&	$\delta$	&	[Zn/Fe]	&	$\delta$	&	[Ba/Fe]	&	$\delta$	&	[La/Fe]	&	$\delta$	&	[Nd/Fe]	&	$\delta$	&	[Eu/Fe]	&	$\delta$	\\
ET0009	&$	-1.68	$&	0.16	&$	 ...	$&	 ...	&$	 ...	$&	 ...	&$	0.57	$&	0.22	&$	 ...	$&	 ...	&$	0.20	$&	0.07	&$	 ...	$&	 ...	&$	0.02	$&	0.11	&$	0.10	$&	0.26	&$	-0.09	$&	0.15	&$	0.06	$&	0.28	&$	0.10	$&	0.18	&$	-0.02	$&	0.15	&$	 ...	$&	 ...	&$	-0.22	$&	0.22	&$	0.15	$&	0.22	&$	 ...	$&	 ...	&$	0.31	$&	0.28	\\
ET0013	&$	-1.68	$&	0.21	&$	 ...	$&	 ...	&$	 ...	$&	 ...	&$	0.53	$&	0.27	&$	 ...	$&	 ...	&$	0.28	$&	0.10	&$	 ...	$&	 ...	&$	0.09	$&	0.18	&$	0.67	$&	0.29	&$	-0.07	$&	0.18	&$	0.13	$&	0.35	&$	 ...	$&	 ...	&$	 ...	$&	 ...	&$	 ...	$&	 ...	&$	-0.29	$&	0.32	&$	 ...	$&	 ...	&$	 ...	$&	 ...	&$	 ...	$&	 ...	\\
ET0024	&$	-1.24	$&	0.10	&$	-0.10	$&	0.20	&$	-0.55	$&	0.21	&$	 ...	$&	 ...	&$	 ...	$&	 ...	&$	0.00	$&	0.14	&$	-0.61	$&	0.22	&$	-0.05	$&	0.14	&$	-0.23	$&	0.23	&$	-0.26	$&	0.22	&$	0.17	$&	0.29	&$	-0.07	$&	0.18	&$	-0.39	$&	0.13	&$	 ...	$&	 ...	&$	 ...	$&	 ...	&$	-0.30	$&	0.16	&$	 ...	$&	 ...	&$	-0.17	$&	0.22	\\
ET0026	&$	-1.80	$&	0.16	&$	 ...	$&	 ...	&$	 ...	$&	 ...	&$	0.44	$&	0.19	&$	 ...	$&	 ...	&$	0.07	$&	0.06	&$	 ...	$&	 ...	&$	-0.09	$&	0.12	&$	0.03	$&	0.22	&$	-0.32	$&	0.15	&$	0.07	$&	0.25	&$	-0.12	$&	0.16	&$	-0.07	$&	0.19	&$	 ...	$&	 ...	&$	-0.24	$&	0.21	&$	 ...	$&	 ...	&$	 ...	$&	 ...	&$	 ...	$&	 ...	\\
\hline							
\end{tabular}
\end{sidewaystable*}
\setcounter{table}{4}
\clearpage
\begin{sidewaystable*}
\caption{(cont’d)\label{table:abund2}}
\vspace{0.2cm}
\tabcolsep=0.03cm
\scriptsize
\end{sidewaystable*}

\end{appendix}

\end{document}